\documentclass[twocolumn,showpacs,preprintnumbers,amsmath,amssymb,superscriptaddress]{revtex4-2}

\usepackage{graphicx}
\usepackage{dcolumn}
\usepackage{bm}
\usepackage{graphics}
\usepackage{subfigure}
\usepackage{graphicx}
\usepackage{epsfig}
\usepackage{epstopdf}
\usepackage{xcolor}
\usepackage{multirow}
\usepackage{mathtools}
\usepackage[colorlinks=true,citecolor=blue]{hyperref}
\hypersetup{colorlinks=true,citecolor=blue,linkcolor=blue,urlcolor=blue}

\allowdisplaybreaks

\begin{document}

\title{Scale and conformal invariance in rotating interacting few-fermion systems}

\author{Viktor Bekassy}
\email{bekassy@chalmers.se}
\affiliation{Department of Microtechnology and Nanoscience (MC2), Chalmers University of Technology, 41296 Gothenburg, Sweden}

\author{Johannes Hofmann}
\email{johannes.hofmann@physics.gu.se}
\affiliation{Department of Physics, Gothenburg University, 41296 Gothenburg, Sweden}
\affiliation{Nordita, Stockholm University and KTH Royal Institute of Technology, 10691 Stockholm, Sweden}

\date{\today}

\begin{abstract}
We show that rotating two-dimensional Fermi gases possess a nonrelativistic scale and conformal invariance at weak but nonzero interactions, where the scale invariance of universal short-range interactions is not yet broken by quantum effects. We demonstrate the symmetry in the excitation spectrum of few-fermion ensembles in a harmonic trap obtained by exact diagonalization. The excitation spectrum is shown to split in a set of primary states and derived excited states that consist of breathing modes as well as two different center-of-mass excitations, which describe cyclotron and guiding-center excitations of the total particle cloud. Furthermore, the conformal symmetry is manifest in the many-body wave function, where it dictates the form of the hyperradial component, which we demonstrate using Monte Carlo sampling of few-body wave functions.
\end{abstract}

\maketitle

\section{Introduction}

Ultracold quantum gas experiments are used to simulate strongly correlated phases of matter, and in particular to create artificial gauge fields to emulate the physics of the lowest Landau level. In its simplest setting, a synthetic magnetic field is induced in a trapped two-dimensional gas brought in rotation~\cite{Madison00,aboshaer01,Zwierlein05,Fletcher21,Fletcher23}, which is described in the rotating frame by a substitution~\cite{bloch08}
\begin{align}\label{eq:rotatinghamiltonian}
H \to H(\Omega) = H - \Omega L_z ,
\end{align}
where $H$ is the many-body Hamiltonian of the nonrotating system, $\Omega$ is the rotation frequency, and $L_z$ is the out-of-plane angular momentum component. 
Here, the Hamiltonian $H$ describes nonrelativistic atoms in a harmonic trap with frequency $\omega$ that interact with a short-range potential of strength~$g$. 
The Coriolis force acting on a particle then takes the same form as the Lorentz force on a unit charge in a constant magnetic field of strength $B = 2 m^* \Omega$ ($m^*$ is the atomic mass), with an additional centrifugal force that weakens the harmonic trap confinement~\cite{bloch08}.
In the limit of fast rotation with a frequency that approaches the trap frequency, the effective trap potential vanishes and single-particle levels form fully degenerate Landau levels. 
While such a rapidly rotating gas in the lowest Landau level is seemingly scale invariant due to the complete quenching of the kinetic energy, and described by a single Haldane pseudopotential parameter, the noncommutative nature of the guiding center coordinates violates such a scaling symmetry and gives rise to a quantum anomaly~\cite{hofmann23}. A different quantum anomaly arises if interactions are sufficiently strong to induce transitions between Landau levels: In this case, the contact interaction is renormalized due to virtual excitations, which has been studied extensively in nonrotating systems~\cite{olshanii10,vogt12,gao12,chafin13,peppler18,holten18,drut18,mulkerin18,daza18,hu19,yin20}. Deviations from scale invariance caused by virtual excitations are experimentally observable in a shift of the breathing mode frequency~\cite{olshanii10,hofmann12,peppler18,holten18}, a logarithmic scaling correction to the rf spectrum~\cite{langmack12}, or the emergence of a finite bulk viscosity~\cite{son07,hofmann20,enss19,nishida19}.
However, as was shown for nonrotating systems in a previous work by the present authors~\cite{bekassy22}, virtual excitations only contribute at second order in the dimensionless interaction strength \mbox{$g/(\ell_{\rm ho}^2\hbar\omega)$} ($\ell_{\rm ho}$ is the harmonic oscillator length), such that the scale symmetry is restored at weak interactions [i.e.,  to linear order ${\it O}(g)$]. In this regime, the scale invariance implies a second symmetry, conformal invariance~\cite{Hagen72,zwerger21}. Since scale transformations do not affect the angular momentum, we expect that this invariance also holds for rotating systems.

\begin{figure*}[t!]
\includegraphics[scale=0.35]{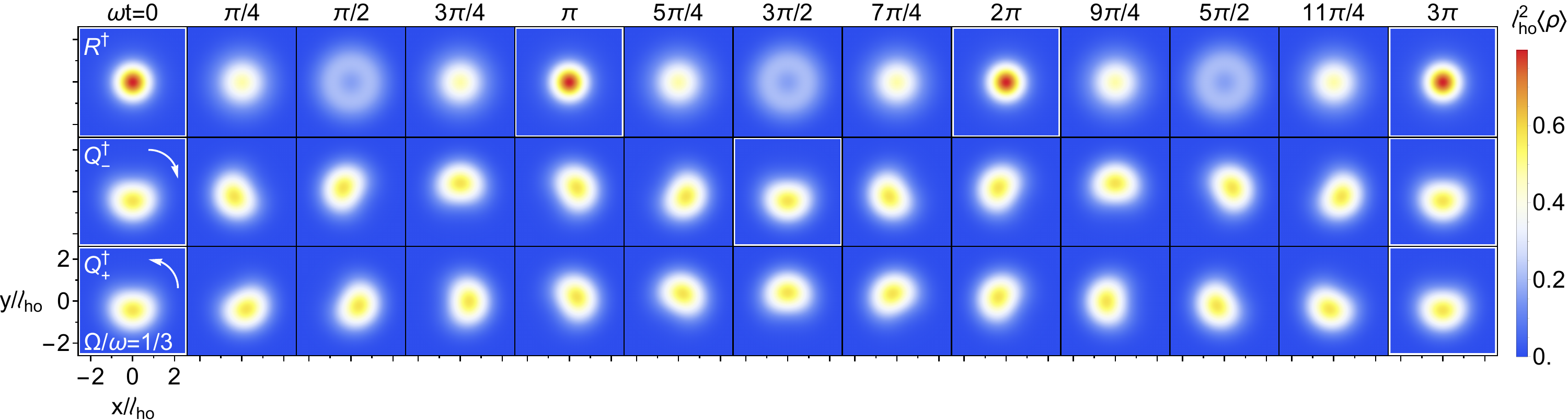}
\caption{Time evolution (left to right panels) of the particle density in the stationary frame with rotation frequency \mbox{$\Omega/\omega=1/3$} for an equal superposition of the \mbox{$N=2$} particle ground state (a primary state) and an excitation by any of the three spectrum-generating operators (top to bottom panels). 
Top panel: The excitation by $R^{\dagger}$ induces an undamped internal breathing mode oscillating at exactly twice the trapping frequency, $2\omega$, independently of both the rotation frequency $\Omega$ and interactions. 
Middle and bottom panels: The center-of-mass excitations $Q_-^{\dagger}$ and $Q_+^{\dagger}$ stir the gas in a clockwise or anticlockwise direction, respectively; the rotation direction is indicated by a white arrow to guide the eye. In contrast to the internal breathing mode excitation, the frequencies of the center-of-mass excitations depend on the rotation, 
$(\omega+\Omega)$ for $Q_-^{\dagger}$, and $(\omega-\Omega)$ for $Q_+^{\dagger}$, corresponding to two complete cyclotron cycles in the middle row and one guiding-center cycle in the bottom row. White frames indicate the oscillation period.
}
\label{fig:1}
\end{figure*}

In this work, we confirm that this is indeed the case and rotating 2D Fermi gases at weak interactions are scale and conformally invariant. We use exact diagonalization and many-body degenerate perturbation theory to reveal signatures of scale invariance in the energy spectrum and the statistics of the many-body wave function. 
A key signature that we establish is that scale and conformal invariance constrain the spectrum of the harmonically trapped rotating gas~\cite{pitaevskii97,castin04,werner06,castin12,zwerger21}, which separates in a set of so-called primary states and their excitations. 
The primary states include the ground state and are specific to the particular system. In particular, they depend on the rotation frequency, such that their energy will change compared to the nonrotating gas~\cite{bekassy22}. From each primary state, we find an infinite set of derived states that are composed of three different excitations: (i) breathing modes, (ii) cyclotron center-of-mass excitations, and (iii) guiding-center center-of-mass  excitations. The breathing modes are constrained by the conformal symmetry to an excitation energy $2\omega$ of exactly twice the trap frequency, independent of interactions, while the center-of-mass excitations follow from Galilean invariance and have excitation energies $\omega+\Omega$ and $\omega-\Omega$, respectively. 
The latter two excitations correspond to a cyclotron motion of the center of mass and a drift of the center-of-mass guiding center, respectively. 
This change in the center-of-mass excitations is a further difference compared to nonrotating systems~\cite{bekassy22}. 
Microscopically, the conformal tower structure follows because the Hamiltonian of a rotating trapped gas is part of a symmetry algebra (specifically, the trap potential is at the same time the generator of special conformal transformations~\cite{mehen00,son06,nishida07}).
From the symmetry algebra, excitation operators can be created, which we denote by $R^\dagger$, $Q_+^\dagger$, and $Q_-^\dagger$ throughout the paper.

\begin{figure}[b!]
\includegraphics[scale=0.62]{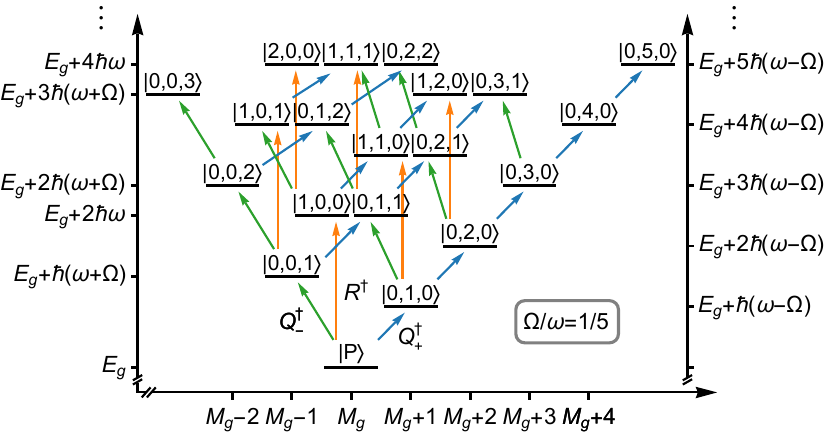}
\caption{
The conformal tower with the first nonprimary states derived from a primary state $|P\rangle$ with energy $E_g$ and total angular momentum $M_g$, shown here for a rotation frequency
$\Omega/\omega=1/5$. Blue arrows are center-of-mass excitations created by $Q_+^{\dagger}$ that increase the energy by \mbox{$\hbar(\omega-\Omega)$} and the angular momentum by $+1$, i.e., that stir in the direction of rotation (see Fig.~\ref{fig:1}). Green arrows are center-of-mass excitations created with $Q_-^{\dagger}$ that increase the energy by \mbox{$\hbar(\omega+\Omega)$} and decrease the angular momentum by $-1$. Orange arrows are internal breathing modes created by $R^{\dagger}$ that increase the energy by $2\hbar\omega$ while preserving the angular momentum.
}
\label{fig:2}
\end{figure}

To illustrate the nature and naming of these excitations, we show in Fig.~\ref{fig:1} density plots in the stationary two-dimensional plane of an equal superposition $|\Psi_0(t)\rangle + |\Psi_e(t)\rangle$ of a ground state wave function $|\Psi_0\rangle$ and the first excited breathing and center-of-mass states $|\Psi_e\rangle = R^\dagger |\Psi_0\rangle, Q_-^\dagger |\Psi_0\rangle$, or $Q_+^\dagger |\Psi_0\rangle$, respectively (top to bottom). These wave functions are obtained using the calculations in this paper for \mbox{$N=2$} weakly interacting particles in a harmonic trap that rotates at a third of the trap frequency, $\Omega/\omega=1/3$. The density of a superposition of eigenstates with different energies evolves in time, and horizontal panels show the density plot at time increments $\Delta t = \pi/(4\omega)$ up to one-and-a-half trap periods \mbox{$T=2\pi/\omega$}. The top panel shows a breathing mode excitation, and indeed the gas is seen to radially expand and contract. As is apparent from the figure, the mode is undamped and completes three cycles in the time period, corresponding to an oscillation frequency of $2\omega$. (We mark the oscillation period by white frames in Fig.~\ref{fig:1} to guide the eye.) The middle panel shows an undamped center-of-mass oscillation---i.e., the atomic cloud moves without any internal deformation---which is seen to complete two full periods in clockwise direction at an increased frequency \mbox{$\omega + \Omega =  4\omega/3$}; this is the analog of classical cyclotron motion. The bottom panel shows a second undamped center-of-mass excitation, which rotates in the counterclockwise direction with reduced frequency $\omega - \Omega =  2\omega/3$, completing one full rotation, and which is the analog of classical guiding-center motion.

In combination, starting from any primary state $|P\rangle$ with energy $E_g$ and angular momentum $M_g$, an infinite set of breathing and center-of-mass excitations is obtained. We illustrate this conformal tower structure in Fig.~\ref{fig:2} (here explicitly for a rotation frequency \mbox{$\Omega/\omega = 1/5$}), where the horizontal axis shows the angular momentum and the vertical axis the excitation energy. Vertical orange arrows denote breathing mode excitations, which do not change the angular momentum; green arrows  cyclotron center-of-mass excitations, which reduce the total angular momentum; and blue arrows guiding-center center-of-mass excitations, which increase the total angular momentum. Every state thus has an associated primary state, which forms the bottom of a conformal tower, and is specified by the number $|a, b, c\rangle$ of breathing and center-of-mass excitations (note that since the excitations are independent, the order in which they are excited is not important).
The conjecture is then that for the full excitation spectrum of the weakly interacting rotating gas, we can identify the primary states and all derived excitations in the conformal tower. Indeed, in this paper we confirm the conformal tower structure in the energy spectrum for few-particle ensembles. In addition, we compute and confirm the hyperradial distribution of the many-body wave function using Metropolis importance sampling. We consider few-fermion ensembles of two-component Fermi gases, and our predictions should be observable in experiments on interacting few-body 2D Fermi systems with recently developed single-particle imaging techniques~\cite{Bergschneider18,bayha20,holten21a,holten21b}.  

This paper is structured as follows: Section~\ref{sec:2} discusses the level structure of two-component Fermi gases in a rotating harmonic trap and introduces degenerate perturbation theory. Section~\ref{sec:primary} then discusses the spectrum-generating conformal symmetry algebra that gives rise to the conformal tower structure shown in Fig.~\ref{fig:2}. These predictions for the level structure are then explicitly verified in our numerical calculations presented in Sec.~\ref{sec:IV}. Additional predictions for the hyperradial part of the many-body wave function are confirmed using Monte Carlo sampling of our eigenstates and presented in Sec.~\ref{sec:hyperradial}. The paper contains two Appendixes with a derivation of the center-of-mass and the hyperradial wave functions starting from the operator algebra as well as details of the Monte Carlo sampling.

\section{Properties of Rotating 2D Fermi Gases}\label{sec:2}

The aim of our work is to reveal the conformal symmetry in the excitation spectrum and many-body wave function for few-fermion ensembles in a rotating harmonic trap with weak contact interactions. This section sets the groundwork for these calculations and discusses the basics of the level structure of rotating Fermi gases, both for free fermions and for contact interactions, and introduces degenerate perturbation theory for weak interactions.

Throughout the paper, we consider two-component fermions with spin projection \mbox{$\sigma=\uparrow,\downarrow$} and mass $m^*$ (we include an asterisk to avoid possible confusion with an angular momentum quantum number) that are confined in a two-dimensional harmonic trap with oscillator frequency $\omega$ and rotation frequency $\Omega$.  We consider fixed-particle number states with \mbox{$N=N_{\uparrow}+N_{\downarrow}$} atoms that contain an equal number of both spin types. We use dimensionless units where both the oscillator energy \mbox{$\hbar \omega = 1$} and the oscillator length \mbox{$\ell_{\rm ho}=\sqrt{\hbar/m^*\omega} = 1$} are set to unity (in particular, the rotation frequency is measured in units of $\omega$). We restore full units in the plots for clarity.

\subsection{Noninteracting rotating Fermi gas}\label{sec:IIA}

The noninteracting dimensionless Hamiltonian in a harmonic rotating trap in the stationary frame is
\begin{equation}\label{eq:hamilton}
H^{(0)}(\Omega) = \sum_{j\sigma} \biggl( - \frac{1}{2} \nabla^2_{j\sigma} + \frac{r_{j\sigma}^2}{2} + \Omega\, i \frac{\partial}{\partial \varphi_{j\sigma}} \biggr) ,
\end{equation}
where $r_{j\sigma}$ and $\varphi_{j\sigma}$ label the position of particle $j$ in polar coordinates. The first term is the kinetic energy, the second term describes the harmonic trap potential, and the last term is the out-of-plane-component of the angular momentum operator. The Hamiltonian may be rewritten with a vector potential \mbox{${\bf A} = m^*{\bf e}_z \times \Omega {\bf r}$}, which describes a unit charged particle in a constant perpendicular magnetic field of strength ${\bf B} =2 m^* \bf{\Omega}$, indicating the mathematical equivalence of the Coriolis force and the magnetic Lorentz force on a charged particle~\cite{bloch08}. In addition, after separating the vector potential, the particles experience a reduced trapping potential $1-\Omega^2$, such that $\Omega \leq 1$ must hold to ensure that the spectrum is bounded, or physically, that the centrifugal force does not overcome the trapping force. 

\begin{figure}[t!]
\includegraphics[scale=0.425]{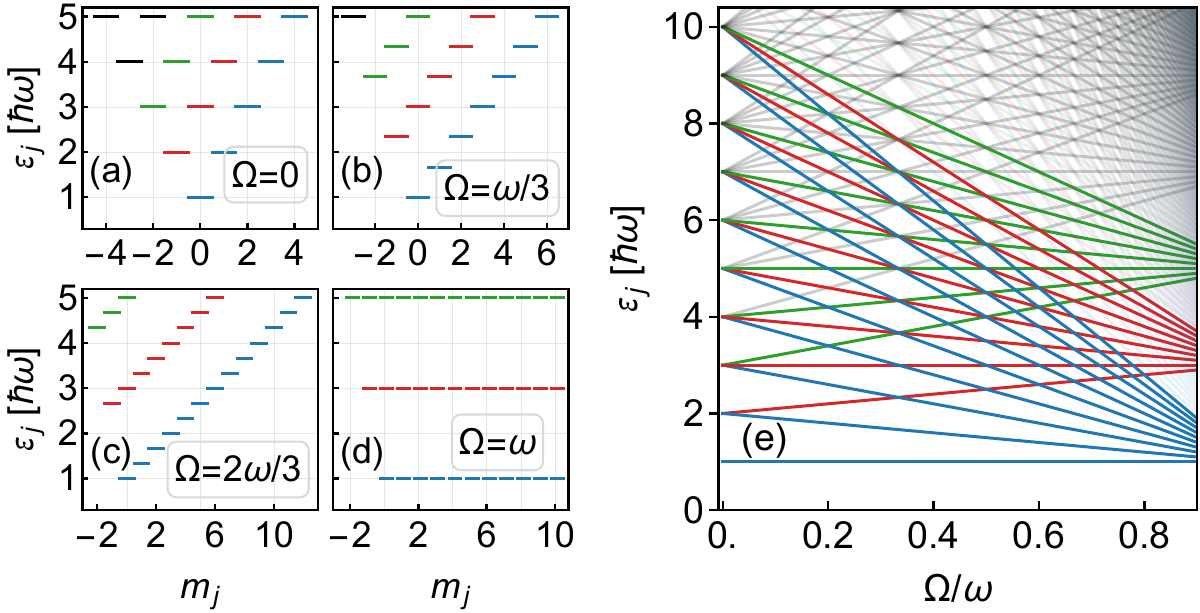}
\caption{
(a)--(d) Single-particle spectrum of particles in a rotating trap ordered by angular momentum for four different rotation frequencies \mbox{$\Omega/\omega=0, 1/3, 2/3,$ and $1$}. States with quantum number \mbox{$n_j=0,1,2$} are highlighted in blue, red, and green, respectively. In (d), the single-particle energy states form Landau levels.
(e) Evolution of the single-particle spectrum as a function of rotation frequency $\Omega$.
}
\label{fig:3}
\end{figure}

Single-particle eigenstates of the Hamiltonian~\eqref{eq:hamilton} are described by two quantum numbers \mbox{$j=\{n_j,k_j\}$} with \mbox{$n_j,k_j \geq 0$} and a harmonic oscillator wave function~\cite{bloch08}
\begin{align}
\phi_j(z,\bar{z}) = \sqrt{\frac{\min(n_j,k_j)!}{\pi \max(n_j,k_j)!}} \ z^{k_j-n_j} e^{-\frac{\bar{z}z}{2}} \, L_{\min(n_j,k_j)}^{|k_j-n_j|}(\bar{z}z) , \label{eq:singleparticlewf}
\end{align}
where $L_{\min(n_j,k_j)}^{|k_j-n_j|}$ is an associated Laguerre polynomial and we use complex coordinates \mbox{$z = r e^{i\varphi}$}. These states are eigenstates of the angular momentum operator with eigenvalue \mbox{$m_j = k_j - n_j \geq - n_j$}.
The corresponding eigenenergies are
\begin{equation}\label{eq:ensp}
\epsilon_j=1 + (1+\Omega) n_j + (1-\Omega) k_j . 
\end{equation} 
Without rotation (\mbox{$\Omega=0$}), this is the spectrum of the two-dimensional harmonic oscillator, where energy levels with energy \mbox{$\ell+1$} are \mbox{$\ell+1$}-fold degenerate with degenerate states distinguished by their angular momentum projection \mbox{$m_j=-\ell,-\ell+2,\ldots,\ell$} (corresponding to \mbox{$n_j = 0,1,\ldots,\ell$}). This is illustrated in Fig.~\ref{fig:3}(a), where states with \mbox{$n_j=0,1,2$} are marked in blue, red, and green, respectively. In a rotating trap [cf. Figs.~\ref{fig:3}(b) and (c) for two rotation frequencies $\Omega=1/3$ and $2/3$], the single-particle levels shift by an amount set by their angular momentum: For angular momenta along the direction of rotation (positive  $m_j$), the energy decreases by $m_j \Omega$; for negative angular momenta, the energy increases by the same amount. 
 As is apparent from the figures, new level degeneracies arise with changing rotation frequency.
Finally, in the limit $\Omega\rightarrow1^{-}$ where the fermions are no longer trapped [Fig.~\ref{fig:3}(d)], states with fixed $n_j$ form degenerate Landau levels that are separated by $2\Omega$. 
The full evolution of the single-particle spectrum (without resolving the angular momentum) is illustrated in Fig.~\ref{fig:3}(e), where new degeneracies are visible at rational fractions $\Omega  = p/q$ with $p, q \in \mathbb{N}$. 

A noninteracting few-particle eigenstate $|\Phi \rangle$ is described by a set of occupied single-particle levels $\{\lambda_1, \lambda_2,\ldots\}$, where each level accommodates at most one particle of each spin type~\cite{VBthesis}. In a position-space projection, these states are represented as a Slater determinant of the single-particle wave functions in Eq.~\eqref{eq:singleparticlewf} as~\cite{slater1929,slater1931,condon1930,lowdin55} 
\begin{align} \label{eq:slater_wf}
\langle \mathbf{r}_{1\uparrow}, \ldots, \mathbf{r}_{1\downarrow},\ldots |\Phi \rangle=\Phi_{\uparrow}( \mathbf{r}_{1\uparrow}, \ldots) \cdot \Phi_{\downarrow}( \mathbf{r}_{1\downarrow}, \ldots)
\end{align}
with
\begin{align} \label{eq:slater}
\begin{split}
&\Phi_{\sigma}( \mathbf{r}_{1\sigma}, \ldots, \mathbf{r}_{N_{\sigma}\sigma})     \\[1ex]
&= \frac{1}{\sqrt{N_{\sigma} !}} \left|
\begin{array}{ccc}
\phi_{\lambda_1}(\mathbf{r}_{1\sigma})  &
\cdots & \phi_{\lambda_{N_\sigma}}(\mathbf{r}_{1\sigma}) \\[1ex]
\vdots & \ddots & \vdots \\[1ex]
\phi_{\lambda_1}(\mathbf{r}_{{N_{\sigma}}\sigma}) &\cdots & \phi_{\lambda_{N_{\sigma}} }(\mathbf{r}_{{N_{\sigma}}\sigma})
\end{array}\right| .
\end{split}
\end{align}
Such basis states are odd under any exchange of the $N_\uparrow$ positions $\{\mathbf{r}_{1\uparrow},\ldots\}$ or the $N_\downarrow$ positions $\{\mathbf{r}_{1\downarrow},\ldots\}$,  reflecting the Pauli principle. 
Energy eigenstates in a rotating isotropic trap are also simultaneous total angular momentum eigenstates with
\begin{align}
M &= \sum_{j=1}^{N_\uparrow} m_{\lambda_j} + \sum_{j=1}^{N_\downarrow} m_{\lambda_j} ,
\end{align}
which is the sum of the angular momentum projections of occupied single-particle states.  

\begin{figure}[t!]
\includegraphics[scale=0.425]{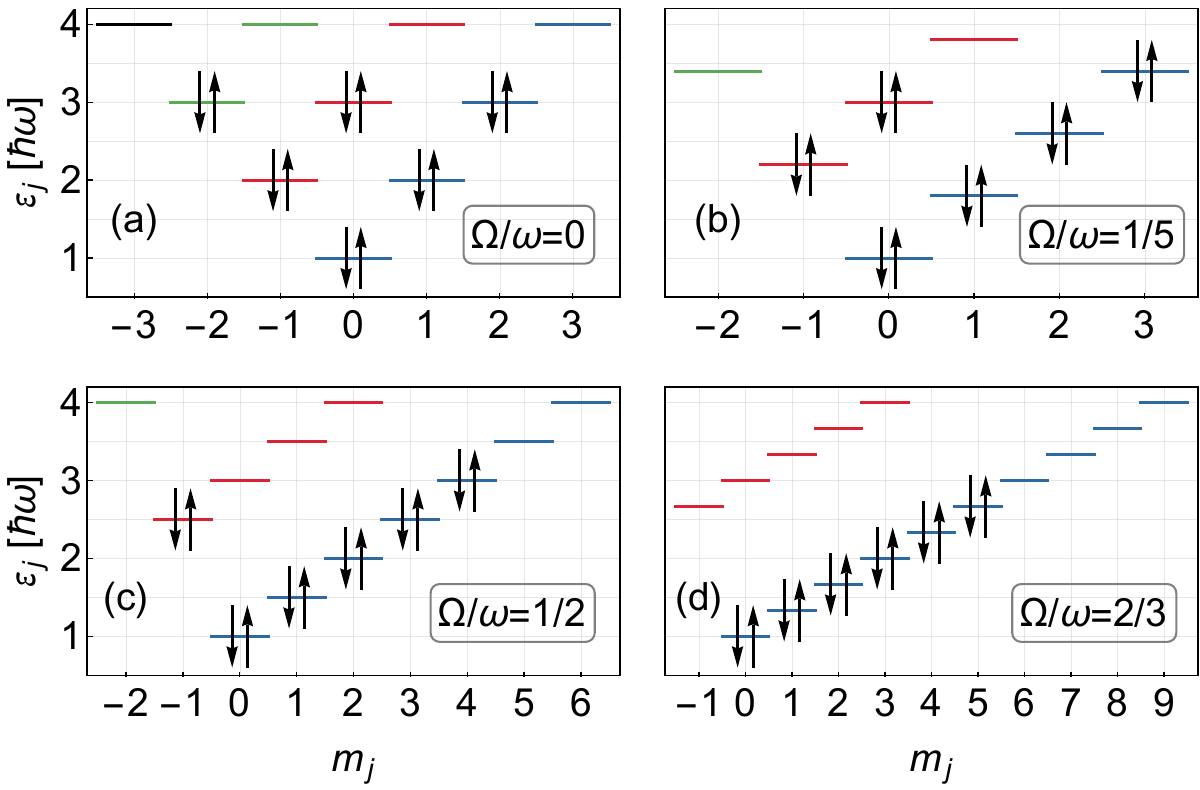}
\caption{(a) Ground state of \mbox{$N=12$} particles in a nonrotating trap. (b)--(d) Threshold rotation frequencies at which the ground state rearranges to a state with larger total angular momentum. The color coding is the same as in Fig.~\ref{fig:3}.
}
\label{fig:4}
\end{figure}

The ground state configuration at a given rotation frequency is obtained by successively populating the lowest single-particle levels with both spins. States obtained in this way are degenerate unless all states at the valence level (the occupied level with highest single-particle energy) are fully occupied. Without rotation, such nondegenerate ground states exist for the ``magic'' numbers $N=2,6,12,20,30,42,\dots$ with completely filled shells [cf. Fig.~\ref{fig:4}(a) for the case \mbox{$N=12$}]~\cite{resare22}. As $\Omega$ increases and the single-particle spectrum changes, new degeneracies emerge and the ground state will change in favor of a state with higher total angular momentum. To illustrate this, we show in Figs.~\ref{fig:4}(b)--4(d) the ground state occupancy at the threshold frequencies $\Omega=1/5,1/2,$ and $2/3$, where the state is degenerate with states with smaller total angular momentum. 
 
\begin{figure}[t!]
\includegraphics[scale=0.425]{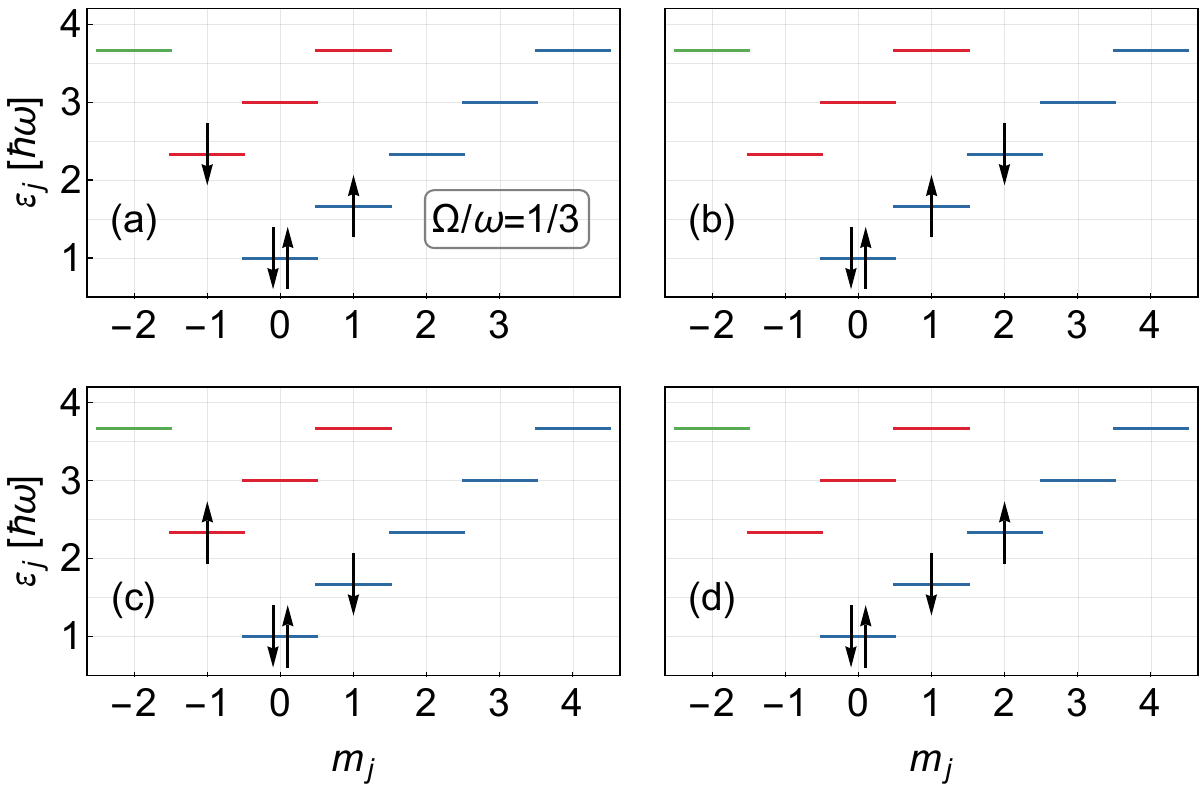}
\caption{Fourfold degenerate excited states for the lowest excitation of the \mbox{$N=4$} particle ground state with rotation frequency \mbox{$\Omega/\omega=1/3$}, which has excitation energy $2\hbar \omega/3$. The color coding is the same as in Fig.~\ref{fig:3}.
}
\label{fig:5}
\end{figure}

Excited states with a given fermion number transfer single fermions or pairs from occupied levels to higher single-particle states. Note that, in general, excited states are highly degenerate even if the ground state is not. To illustrate the degeneracy structure, we show in Fig.~\ref{fig:5} the occupancy of the lowest excitation of an \mbox{$N=4$} state with rotation frequency \mbox{$\Omega = 1/3$}. The four excited states have degenerate excitation energy $2/3$. While the number of degenerate states is small in this example, it generally grows very quickly with both particle number and excitation energy. For example, excited states  with excitation energy $2$ for \mbox{$N=12$} particles with \mbox{$\Omega=0$} are $226$-fold degenerate, and for \mbox{$N=20$} particles with \mbox{$\Omega=1/3$} they are $2060$-fold degenerate. In our work, we identify ground and excited state configurations by numerical counting. We emphasize that the complexity of the subspace of degenerate excited states is still vastly smaller than the size of the full Hilbert space for $N$ particles.

\subsection{Contact interactions}\label{sec:IIB}

The ground and excited state degeneracy is lifted when interactions are taken into account. For fermionic quantum gases, these are dominantly short-range $s$-wave interactions between different spin species that are described in a universal way by a delta function potential
\begin{equation}\label{eq:hamiltonint}
H^{({\rm I})} = g \sum_{ij} \delta^{(2)}({\bf r}_{i\uparrow} - {\bf r}_{j\downarrow}),
\end{equation}
with a dimensionless interaction strength $g$. Since the delta function is a homogeneous function under coordinate rescaling \mbox{$\delta^{(2)}(\lambda {\bf r}) = \lambda^{-2} \delta^{(2)}({\bf r})$}, a rescaling of every particle coordinate $\bf r \rightarrow \lambda \bf r$ by a constant $\lambda$ changes the interacting Hamiltonian in the absence of a trapping potential as $H \to H/\lambda^2$; i.e., the kinetic and the interaction energies transform in the same way and the Hamiltonian is classically scale invariant. 
However, a delta function interaction in 2D requires renormalization such that the coupling~$g$ picks up an additional regulator scale that breaks the scale invariance of the interaction~\cite{bloch08}, which is known as a quantum scale anomaly~\cite{olshanii10,hofmann12}. Yet, as argued in~\cite{bekassy22}, we expect that renormalization effects are negligible for weak interactions. 
The coupling is then indeed scale invariant and given by \mbox{$g = \sqrt{8\pi}a_{3D}/l_z$}, with $a_{3D}$  the 3D scattering length and $\ell_z$ the oscillator length of a transverse harmonic potential~\cite{bloch08,zwerger16,hofmann21}.
Hence, to linear order in the interaction strength $g$, the scale invariance of the theory is exact; experimental signatures of the quantum scale anomaly enter only at quadratic order~\cite{hofmann12,langmack12,hofmann20,enss19,nishida19}. 

We therefore use first-order degenerate perturbation theory in $g$ and restrict our attention to few-particle ensembles to stay in a quasi-2D regime where particles only occupy the lowest state of a transverse harmonic
potential. Within first-order degenerate perturbation theory, we collect all states $\{|\Psi_n\rangle\}$ with equal noninteracting energy $E^{(0)}_N$ and diagonalize the Hamiltonian matrix~\cite{sakurai94,gottfried03} 
\begin{equation} \label{eq:matrixelement}
H_{mn} = \langle \Psi_m | H(\Omega) | \Psi_n \rangle 
\end{equation}
to obtain the energy eigenvalues \mbox{$E_N=E_N^{(0)}+E_N^{(1)}$}. Here, \mbox{$E_N^{(1)} \sim {\it O}(g)$} and scale invariance follows directly from the homogeneity
of the delta potential in the matrix element in Eq.~\eqref{eq:matrixelement}. 
Formally, leading-order degenerate perturbation theory will apply for interaction strengths \mbox{$g \ll 1$}, which do not connect many-body states with equal total angular momentum at different noninteracting energies (i.e., the splitting of degenerate states is small compared to the harmonic oscillator spacing). Note that for a fixed particle number $N$, this includes the limit of rapid rotations \mbox{$\Omega\to 1^-$}. Here, degenerate perturbation theory is equivalent to an exact diagonalization in the lowest Landau level~\cite{hofmann23,palm20}.

By definition of a degenerate subspace, the noninteracting contribution to $H_{mn}$ is a diagonal matrix with equal entries $E_N^{(0)}$, which means that the eigenvectors themselves (unlike the eigenenergies) do not depend on the interaction strength $g$. 
Hence, although the eigenvectors we obtain are independent of $g$, they are still a nontrivial superposition of basis states \eqref{eq:slater_wf} governed by the nonrelativistic conformal symmetry. 
Including ${\it O}(g)$ corrections to the eigenvectors corresponds to the next-to-leading order in perturbation theory and involves a divergent summation over all excited states~\cite{sakurai94,gottfried03}, where contributions to eigenenergies are of order ${\it O}(g^2)$. Here, in principle, we anticipate the quantum anomaly to become apparent and the conformal window to close. However, corrections to scale invariance at higher orders can be small~\cite{taylor12}, and we expect the conformal window to extend beyond the range of validity of leading-order perturbation theory. 
Note that another quantum anomaly arises in the rapid rotation limit due to the noncommutative nature of guiding-center coordinates breaking scale invariance~\cite{hofmann23}. 

To evaluate the matrix elements~\eqref{eq:matrixelement}, it is convenient to work in an occupation number representation, in which the Hamiltonian \eqref{eq:rotatinghamiltonian} takes the form
\begin{equation} \label{eq:Ham}
H(\Omega) = \sum_{j,\sigma}\epsilon_{j}c^{\dagger}_{j\sigma}c_{j\sigma}^{} + g \sum_{ijkl}w_{ijkl}c^{\dagger}_{i\uparrow}c^{\dagger}_{j\downarrow}c_{k\downarrow}^{}c_{l\uparrow}^{}.
\end{equation}
Here, $c_{j\sigma}^{\dagger}$ creates a fermion with spin projection \mbox{$\sigma=\uparrow,\downarrow$} in a single-particle state \mbox{$j=\{n_j,k_j\}$} with energy $\varepsilon_j$ given in Eq.~\eqref{eq:ensp}. The interaction matrix element in Eq.~\eqref{eq:Ham} is set by the overlap integral 
\begin{align} \label{eq:overlap}
w_{ijkl}=\int d^2r \, \phi_i^{*}\phi_j^{*}\phi_k\phi_l ,
\end{align}
where $\phi_i$ is the single-particle wave function in Eq.~\eqref{eq:singleparticlewf}. The overlap integral conserves angular momentum (since \mbox{$w_{ijkl}\sim \int_0^{2\pi}\,d\varphi e^{i\varphi(-m_i-m_j+m_k+m_l)} $}), making the choice of single-particle eigenstates~\eqref{eq:singleparticlewf} convenient. 
 
Note that a comprehensive discussion of the ground state properties in a rotating Fermi gas was given by~\textcite{MASHKEVICH07,mashkevich11} in the case of rapid rotations with $\Omega<1$, i.e., involving occupied Landau level states with $n_j=0$. Here, the analyticity of the many-body wave function allows for an exact calculation of the ground state energy even for a general pairwise interaction potential, not just a contact interaction. While excited states within the lowest Landau level can be evaluated in principle using the same method~\cite{mashkevich11}, such excitations do not include the breathing mode excitations, which connect different Landau levels, as will be discussed in the next section. 

\section{Primary States and Conformal Towers}\label{sec:primary}

In this section, we derive in detail the decomposition of the excitation spectrum into  conformal towers composed of primary states and their center-of-mass as well as internal breathing mode excitations, which is illustrated in Fig.~\ref{fig:2}.  
The starting point is a spectrum-generating operator implied by the nonrelativistic conformal symmetry~\cite{pitaevskii97,castin04,werner06,castin12,zwerger21,bekassy22}
\begin{equation} \label{eq:defL}
L^\dagger=  i D + H - 2C ,
\end{equation}
where $H$ is the interacting Hamiltonian without rotation, cf. Eq.~\eqref{eq:rotatinghamiltonian}, $C$ is the generator of special conformal transformations $(t,{\bf r}) \to (t,{\bf r})/(1+\lambda t)$, and $D$ is the generator of scale transformations $(t,{\bf r}) \to (t/\lambda^2,{\bf r}/\lambda)$. 
The commutators $[H(\Omega), L^{\dagger}]=2L^{\dagger}$ and  $[L_z, L^{\dagger}]=0$ imply that when acting on an energy eigenstate, $L^\dagger$ creates an excitation at exactly twice the trapping frequency without any change in the angular momentum. This is also evident in an occupation number representation, where (to leading order in perturbation theory) $L^{\dagger}$ is a single-particle operator
\begin{equation} \label{eq:L_occ}
L^{\dagger}= \sum_{i,j}
\bigl[ 2\sqrt{n_ik_i} \delta_{(n_i,k_i),(n_j+1,k_j+1)}\bigr]
\sum_{\sigma}c^{\dagger}_{i\sigma} c_{j\sigma}
\end{equation}
that creates excitations from a state \mbox{$\{n_j, k_j\}$} to \mbox{$\{n_j+1,k_j+1\}$}. 
Since all operators in Eq.~\eqref{eq:defL} commute with the angular momentum operator, these results continue to hold in a rotating trap. As discussed in the introduction, the excitation is interpreted as an undamped breathing mode excitation. However, it is important to note that the operator $L^\dagger$ mixes internal motion and center-of-mass motion, as we discuss in the following. 

\subsection{Center-of-mass excitations}

In order to demonstrate and disentangle the mixing of internal and center-of-mass excitations, we introduce two additional independent spectrum-generating operators 

\begin{align} \label{eq:defQ_Z1}
Q_+^\dagger &= \frac{i}{\sqrt{4N}} \biggl( N Z  - 2 \frac{\partial}{\partial \bar{Z}} \biggr) , \\
Q_-^\dagger &= \frac{i}{\sqrt{4N}} \biggl(N \bar{Z}  - 2 \frac{\partial}{\partial Z} \biggr) , \label{eq:defQ_Z2}
\end{align}
which depend on the center-of-mass coordinate \mbox{$Z=(1/N)\sum_{i\sigma} z_{i\sigma}$}.  The center-of-mass excitations generated by $Q_\pm^\dagger$ are illustrated in Fig.~\ref{fig:1}. 
These operators have a simple interpretation: They create cyclotron and guiding-center excitations, respectively, for a particle with mass $Nm^*$ in an effective magnetic field \mbox{$B=2Nm^*\Omega$}~\cite{macdonald94}.
They obey the commutation relations \mbox{$[Q_{\pm},Q_{\pm}^\dagger]=1$},  \mbox{$[H(\Omega), Q^{\dagger}_{\pm}]=(1\mp \Omega) Q^{\dagger}_{\pm}$}, and  \mbox{$[L_z,Q_{\pm}^{\dagger}]=\pm Q_{\pm}^{\dagger}$}, which implies that $Q_+^\dagger$ creates an excitation with energy \mbox{$1-\Omega$} and increases the angular momentum by one unit, and \mbox{$Q_-^\dagger$} has excitation energy \mbox{$1+\Omega$} and decreases the angular momentum, where the change in angular momentum is indicated by the subscript. These results are completely independent of interactions, which only affect internal degrees of freedom. In the limit of fast rotations \mbox{$\Omega\to1^-$}, the operator $Q_-^\dagger$ generates the cyclotron resonance between different Landau levels with fixed excitation energy $2\Omega$~\cite{kohn61}, while the operator $Q_+^\dagger$ generates gapless excitations that decrease the filling fraction. Note that in the lowest Landau level limit of rapid rotations, the guiding-center excitation by $Q_+^\dagger$ corresponds to a quasihole excitation~\cite{kasner02}.

The nature of the center-of-mass excitations also becomes clear in an occupation number representation: The (single-particle) operator $Q_-^\dagger$ creates excitations from an occupied state \mbox{$\{n_j, k_j\}$} to a state with a higher Landau-level index \mbox{$\{n_j+1,k_j\}$}, 
\begin{equation} \label{eq:Qm_occ}
Q^{\dagger}_-=\frac{1}{\sqrt{2N}}\sum_{i,j}\bigl[(-1)^{p_-} i\sqrt{2n_i} \delta_{(n_i,k_i),(n_j+1,k_j)}\bigr]
\, \sum_{\sigma} c^{\dagger}_{i\sigma} c_{j\sigma},
\end{equation}
where $p_-=0$ if $n_i>k_i$ and $p_-=1$ if $n_i \leq k_i$, 
whereas $Q_+^\dagger$ excites to levels \mbox{$\{n_j,k_j+1\}$} without changing the Landau level
\begin{equation} \label{eq:Qp_occ}
Q^{\dagger}_+=\frac{1}{\sqrt{2N}}\sum_{i,j}
\bigl[(-1)^{p_+} i\sqrt{2k_i} \delta_{(n_i,k_i),(n_j,k_j+1)}\bigr]\sum_{\sigma}c^{\dagger}_{i\sigma} c_{j\sigma},
\end{equation}
where $p_+=0$ if $k_i>n_i$ and $p_+=1$ if $k_i \leq n_i$.

\begin{figure}[t!]
\includegraphics[scale=0.425]{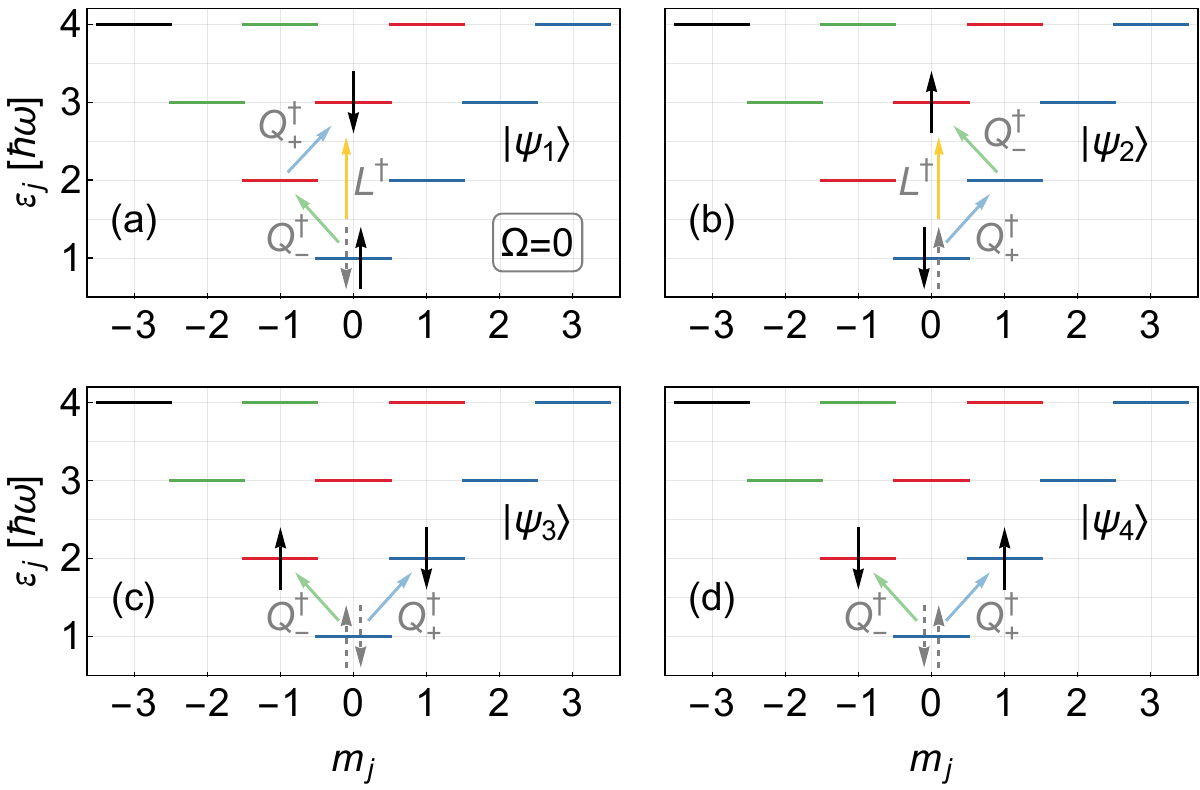}
\caption{
Occupation number representation of the excited states generated by acting with $L^{\dagger}$ [(a) and (b)] and $Q_{+}^{\dagger}Q_{-}^{\dagger}$ [(a)--(d)] on the ground state of $N=2$ particles without rotation ($\Omega=0$). Black (gray) spins indicate occupied excited (ground) single-particle states. We denote the action of $L^{\dagger}$ [Eq.~\eqref{eq:L_occ}] by a yellow arrow, of $Q_{-}^{\dagger}$ [Eq.~\eqref{eq:Qm_occ}] by a green arrow, and of $Q_{+}^{\dagger}$ [Eq.~\eqref{eq:Qp_occ}] by a blue arrow.}
\label{fig:6}
\end{figure}

Returning to the breathing mode excitations, the operators $Q_{\pm}^{\dagger}$ and $L^{\dagger}$ are linearly independent but they do not commute, as can be seen from \mbox{$[L^\dagger, Q_\pm] = - 2 Q_\mp^\dagger$} and \mbox{$[L, Q_\pm^\dagger] = 2 Q_\mp$}. Hence, states generated by $L^\dagger$ and $Q_+^\dagger Q_-^\dagger$ are not orthogonal, which is precisely the statement that a breathing mode generated by $L^\dagger$ also contain center-of-mass excitations.
We illustrate this statement in Fig.~\ref{fig:6} in the occupation number representation for the simple case of the \mbox{$N=2$} ground state $|\text{gs}\rangle$: The operator $L^\dagger$, Eq.~\eqref{eq:L_occ}, creates an equal superposition of two spin states excited from \mbox{$\{n_j, k_j\}$} to \mbox{$\{n_j+1,k_j+1\}$}, \mbox{$L^\dagger|\text{gs}\rangle=2(|\psi_1\rangle+|\psi_2\rangle)$}, where the occupation of the states $|\psi_1\rangle$ and $|\psi_2\rangle$ is illustrated in Figs.~\ref{fig:6}(a) and~(b). The center-of-mass excitation generated by the combination $Q_+^\dagger Q_-^\dagger$ creates the same superposition with additional two-particle excitations, \mbox{$Q_+^\dagger Q_-^\dagger|\text{gs}\rangle=\frac{1}{2}( |\psi_1\rangle+|\psi_2\rangle-|\psi_3\rangle-|\psi_4\rangle )$} [Figs.~\ref{fig:6}(c) and~(d)], which obviously are not orthogonal.

\subsection{Internal breathing mode excitations}

We separate the center-of-mass motion from the bare breathing mode excitation by introducing a spectrum-generating operator of {\it internal} breathing modes
\begin{equation} \label{eq:defR}
R^\dagger = L^\dagger -  \bigl( Q_{+}^{\dagger}Q_{-}^{\dagger} + Q_{-}^{\dagger}Q_{+}^{\dagger} \bigr) ,
\end{equation} 
which commutes with the center-of-mass operators $Q_\pm^\dagger$. This form of the operator is the same as for a nonrotating trap~\cite{werner06,castin12,moroz12,bekassy22}.
Furthermore, it obeys the same commutation relations as $L^\dagger$, \mbox{$[H({\Omega}),R^{\dagger}]=2R^{\dagger}$} and \mbox{$[L_z,R^{\dagger}]=0$}, such that $R^{\dagger}$ creates an excitation with energy $2$ without changing angular momentum. The effect of the operator $R^{\dagger}$ is to generate the internal breathing mode as illustrated in Fig.~\ref{fig:1} in the introduction.  It can be shown that $R^\dagger$ acts on the internal hyperradius 
\begin{align} \label{eq:hypradius}
\tilde{R} = \sqrt{\sum_{i\sigma} |z_{i\sigma} - Z|^2} ,
\end{align}
which gives the coordinate representation of the internal breathing mode operator
\begin{align} \label{eq:R_rep}
R^{\dagger}  &= (N-1) +\tilde{R}\frac{\partial}{\partial \tilde{R}} +s+1+2a -\tilde{R}^2 ,
\end{align}
where $s+1+2a$ parametrizes the internal energy of an eigenstate in a nonrotating trap, and $a$, $s$ are defined in the following section in Eqs.~\eqref{eq:energy_abc} and~\eqref{eq:Eg_s}, respectively. 

In an occupation number representation, states excited by $\tilde{R}^\dagger$ now contain additional two-particle excitations such that they are orthogonal to the center-of-mass excitation $Q_+^\dagger Q_-^\dagger$. In the example in Fig.~\ref{fig:6}, we have \mbox{$R^\dagger|\text{gs}\rangle=|\psi_1\rangle+|\psi_2\rangle+|\psi_3\rangle+|\psi_4\rangle$}, which contains the breathing mode excitation $L^\dagger|\text{gs}\rangle$ but is now orthogonal to $Q_+^\dagger Q_-^\dagger|\text{gs}\rangle$. 
Note that since $Q_\pm^\dagger$ is of order~${\it O}(N^{-1/2})$, the contribution of $Q_+^{\dagger}Q_-^{\dagger}$ scales as~$\sim 1/N$, and one could naively expect single-particle breathing mode excitations \mbox{$\{n_j, k_j\}$} to \mbox{$\{n_j+1,k_j+1\}$} to dominate for increasing $N$. However, there is also an enhancement of order ${\it O}(N)$ of states accessible by two-particle excitations compared to the breathing mode excitations, such that the relative importance of single- and two-particle excitations should thus remain unchanged as $N$ increases. Interestingly, however, the Pauli principle excludes most two-particle excitations for low-lying energy eigenstates such that  $R^{\dagger}$ is predominantly a single-particle operator for increasing $N$: In a nonrotating trap, for example,  the \mbox{$N=2$} state $R^\dagger|\text{gs}\rangle$ shown in Fig. \ref{fig:6} contains  $50\%$ overlap with the breathing mode excitation, which increases to $90\%$ for $N=6$, to $96\%$ for \mbox{$N=12$}, and to $98\%$ for \mbox{$N=20$}. For even higher energy eigenstates and higher breathing mode excitations, two-particle excitations gain importance again. 

\subsection{Conformal tower structure}

We now discuss the full conformal tower structure shown in Fig.~\ref{fig:2}. Define a primary state $|P\rangle$ that is annihilated by all spectrum-generating operators $R$ and $Q_{\pm}$,
\begin{equation}
R|P\rangle =Q_{+}|P\rangle =Q_-|P\rangle =0.
\end{equation}
Note that the ground state for any $N$ and $\Omega$ is a primary state, but a primary state is not necessarily the ground state: Indeed, there is an infinite number of such states. 

A primary state forms the ground step of a conformal tower of orthogonal excited states (the ``nonprimary'' states) that are created by successively acting on $|P\rangle$ with $R^{\dagger}$ and $Q_{\pm}^{\dagger}$. We denote these states by
\begin{equation} \label{eq:abc}
|a,b,c\rangle_P = (R^\dagger)^a (Q_+^\dagger)^b (Q_-^\dagger)^c |P\rangle. 
\end{equation}
This is the structure illustrated in Fig.~\ref{fig:2}, where the energy and angular momentum of a primary state are denoted by $E_g$ and $M_g$, respectively. Excited states in the figure have energy 
\begin{align} \label{eq:energy_abc}
E_{a,b,c}=E_g +2a +\bigl(1-\Omega\bigr)b+\bigl(1+\Omega\bigr)c
\end{align}
and angular momentum 
\begin{align}
M_{a,b,c}=M_g+b-c ,
\end{align}
while the total spin $S_N$, which defines the eigenvalue $S_N(S_N+1)$ of the operator
\begin{equation} \label{eq:spin}
S^2=\sum_{\sigma , \sigma ^{\prime}}\sum_{i,j} \mathbf{S}_{i\sigma} \cdot \mathbf{S}_{j\sigma^{\prime}}  ,
\end{equation}
where ${\bf S}_{i\sigma} = \frac{1}{2}\boldsymbol{\sigma}$ is the vector of Pauli matrices, is conserved. 
The coefficient $s$ that enters the coordinate representation of the internal breathing mode operator in Eq.~\eqref{eq:R_rep} is defined as
\begin{equation} \label{eq:Eg_s}
E_g=2+s-\Omega M_g ,
\end{equation}
and thus sets the ground step energy in a nonrotating trap ($\Omega=0$). The set of all conformal towers, one for every primary state, forms a complete basis of the Hilbert space. 

\subsection{Casimir operator}

It is further instructive to discuss the separation of internal and center-of-mass motion on a Hamiltonian level:
Introducing internal particle coordinates relative to the center of mass 
\mbox{$\tilde{z}_j=z_j-Z$}
, the Hamiltonian splits into an internal part and a center-of-mass part,
\begin{align}
H(\Omega)=H^{\text{com}}(\Omega)+H^{\text{int}}(\Omega) ,
\end{align}
which always holds for a Galilean invariant interaction. The center-of-mass part describes a fictitious particle of mass $N m^*$ in a rotating harmonic trap and is expressed solely in terms of the operators $Q_{\pm}$:
\begin{align} \label{eq:Hamcm}
\begin{split}
H^{\text{com}}(\Omega) &=1 + \bigl(1-\Omega\bigr) Q_+^{\dagger}Q_+ + \bigl(1+\Omega\bigr) Q_-^{\dagger}Q_-,\\[2ex]
L_z^{\text{com}} & =Q_+^{\dagger}Q_+-Q_-^{\dagger}Q_- .
\end{split}
\end{align}
The decomposition into independent guiding-center and cyclotron excitations of the center of mass is directly visible in this representation. 
For a given excited nonprimary state, the center-of-mass contributions to the energy and angular momentum are
\begin{equation}
\begin{split}
E_{a,b,c}^{\text{com}}&=1+\bigl(1-\Omega\bigr)b+\bigl(1+\Omega\bigr)c, \\[1ex]
M_{a,b,c}^{\text{com}}&=b-c .
\end{split}
\end{equation}
Note that a primary state and its internal breathing mode excitations are completely determined by the relative particle dynamics with internal energy and angular momentum
\begin{equation}
\begin{split}
E_{a,b,c}^{\text{int}}&=E_g-1+2a , \\[1ex]
M_{a,b,c}^{\text{int}}&=M_g .
\end{split}
\end{equation}

In order to disentangle different primary states and their conformal towers, we introduce the SO(2,1) Lie algebra $[T_1, T_2]=-iT_3,\;[T_2, T_3]=iT_1$, and $[T_3, T_1]=iT_2$ with the generators~\cite{bekassy22}
\begin{equation} \label{eq: generators}
\begin{split}
T_1&=\frac{1}{4}\left(R^{\dagger}+R\right) 
 , \\
T_2&=\frac{1}{4i}\left(R^{\dagger}-R\right) 
 , \\
T_3&=\frac{1}{2}\bigl(H - H_{}^{\text{com}}\bigr) =\frac{1}{2}H^{\text{int}} ,
\end{split}
\end{equation}
where $H^{\text{int}}$ and $H^{\text{com}}$ are the internal and center-of-mass parts of the Hamiltonian, respectively, without rotation. This is the algebra of the Lorentz group in 2+1 dimensions, with $T_1$ and $T_2$ generating boosts in two directions, and $T_3$ rotations in the plane~\cite{zwerger16}. 
The Casimir operator of the algebra,
\begin{align}\label{eq:casimir}
T &= 4\left(T_3^2 - T_1^2-T_2^2\right) \nonumber \\
&= \bigl( H^{\text{int}}\bigr)^2-\frac{1}{2}\left(R R^{\dagger}+R^{\dagger}R \right) ,
\end{align}
then commutes with the generators in Eq.~\eqref{eq: generators} and is constant within each conformal tower. Its expectation value is
\begin{equation} \label{eq:exp_cas}
\begin{split}
\langle T \rangle &= \langle a,b,c |T| a,b,c  \rangle_P=(s+1)(s-1),
\end{split}
\end{equation}
with $s$ defined in Eq.~\eqref{eq:Eg_s}.
The value of the Casimir within a conformal tower is thus independent of the rotation frequency. 

\begin{figure*}[t!]
\subfigure{\includegraphics[scale=0.8]{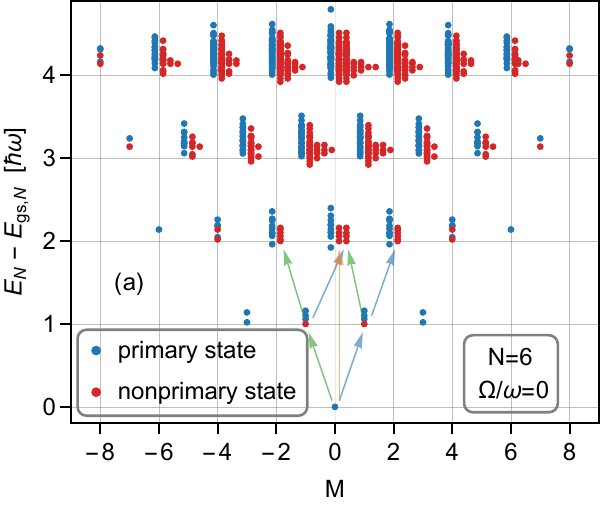}\label{fig:7a}}\qquad
\subfigure{\includegraphics[scale=0.8]{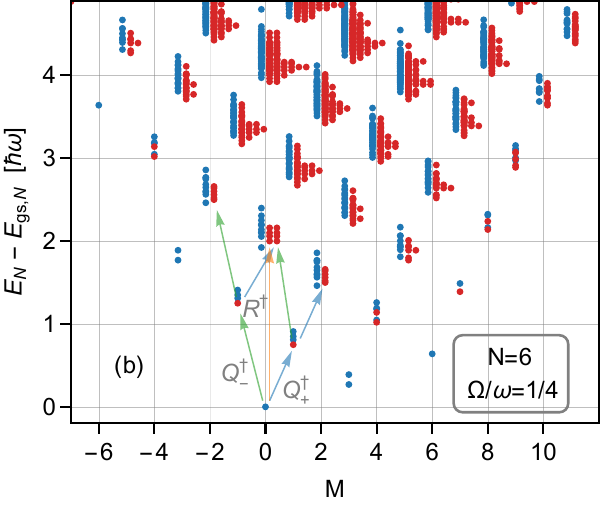}\label{fig:7b}} \\
\subfigure{\includegraphics[scale=0.8]{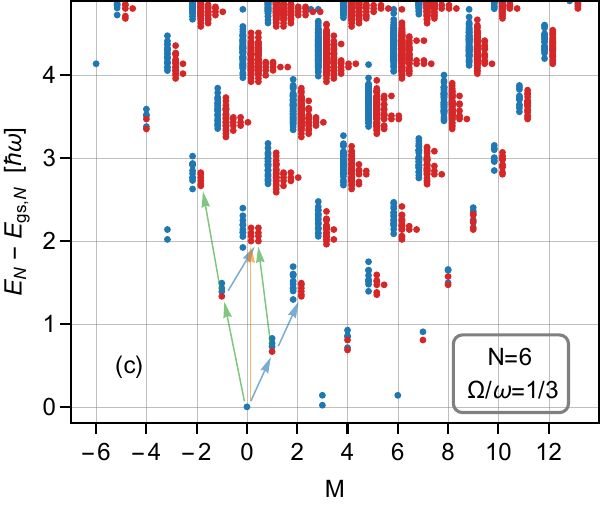}\label{fig:7c}}\qquad
\subfigure{\includegraphics[scale=0.8]{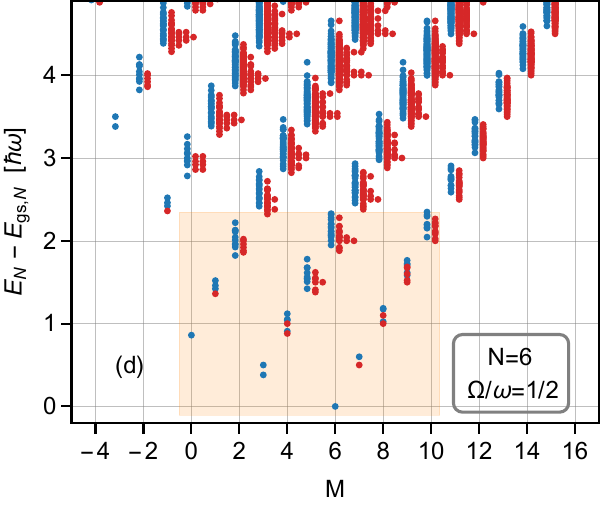}\label{fig:7d}} \qquad
\caption{
Excitation energies for \mbox{$N=6$} particles in a rotating harmonic trap at rotation frequencies \mbox{$\Omega/\omega=0,1/4,1/3$}, and $1/2$, ordered by angular momentum for an attractive interaction \mbox{$g=-1$}. Primary (nonprimary) states are represented by blue (red) points (cf.~Fig.~\ref{fig:2}) and the color coding is consistent in all plots. Overlapping points are moved horizontally for clarity. In (a)--(c), we indicate by arrows the first few states of the lowest conformal tower  originating from the ground state primary state at $M=0$ (compare with Fig.~\ref{fig:2}). Each primary blue state sets the ground step for another conformal tower. Orange-colored region: Lowest 115 energy eigenstates for $N=6$ at $\Omega/\omega=1/2$ used in Sec.~\ref{sec:hyperradial}.
}
\label{fig:7}
\end{figure*}

Following~\cite{werner06,castin12}, we define a ground step operator $H_g(\Omega)$ by inverting Eq.~\eqref{eq:casimir} using $E_{a,b,c}^{\text{int}}=E_g-1$ for primary states (suppressing the dependence on $a,b,c$), and \mbox{$ [R,R^{\dagger}]=4H^{\text{int}}$}, such that
\begin{equation} \label{eq:Hamgs}
H_g(\Omega)=1 + \sqrt{1+T}- \Omega  L_z^{\text{int}},
\end{equation}
where $L_z^{\text{int}}$ is the internal angular momentum and both $H_g(\Omega)$ and $L_z^{\text{int}}$ are constant within a conformal tower. Evaluating the ground step operator for a state yields the internal energy of the primary state of a conformal tower:
\begin{equation}
\begin{split}
H_g(\Omega)|a,b,c\rangle_P &= (E_g-1 )|a,b,c\rangle_P \\
&=(1+s-\Omega M_g)|a,b,c\rangle_P,
\end{split}
\end{equation}
where \mbox{$s=\sqrt{1+\langle T \rangle }$}.
One can then define the rescaled internal breathing mode operator 
\begin{equation}
r^{\dagger }= \frac{1}{\sqrt{2}} R^{\dagger } \Bigl[H^{\text{int}}+H_g(\Omega)+ \Omega L_z^{\text{int}} \Bigr]^{-1/2},
\end{equation}
where now $[r,r^{\dagger}]=1$~\cite{VBthesis}. Thus, the total Hamiltonian and the angular momentum \mbox{$L_z=L_z^{\text{com}}+L_z^{\text{int}}$} are expressed compactly as
\begin{equation}
\begin{split}
H(\Omega)&=H^{\text{com}}(\Omega)+H_g(\Omega)+2r^{\dagger}r , \\[1ex]
L_z&= Q_+^{\dagger}Q_+-Q_-^{\dagger}Q_-+L_z^{\text{int}}.
\end{split}
\end{equation}
In summary, we have established the conformal tower structure in a rotating trap. 
Compared to a nonrotating trap, the effect of rotations is twofold: First, it rearranges primary states through the ground step operator $H_g(\Omega)$, and second, it changes the excitation energy of center-of-mass excitations, yet undamped breathing modes at exactly $2\omega$ remain. 

\section{Conformal structure in few-fermion ensembles}\label{sec:IV}

\begin{figure*}[t!]
\subfigure{\includegraphics[scale=0.8]{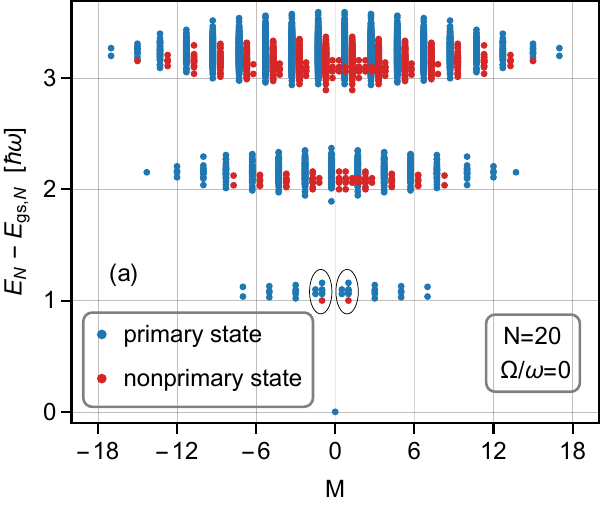}\label{fig:8a}}\qquad
\subfigure{\includegraphics[scale=0.8]{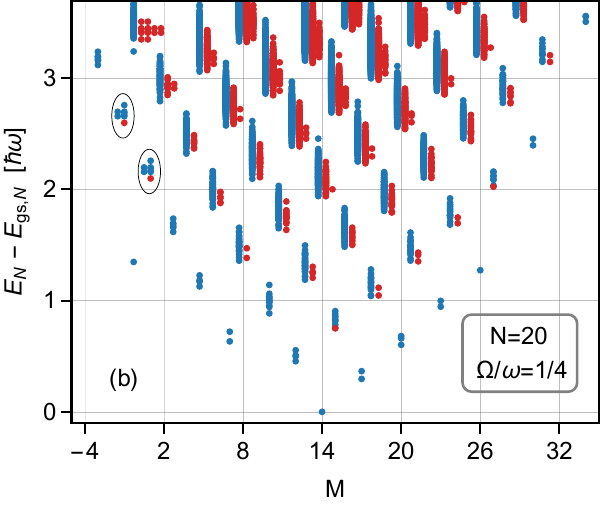}\label{fig:8b}} \\
\subfigure{\includegraphics[scale=0.8]{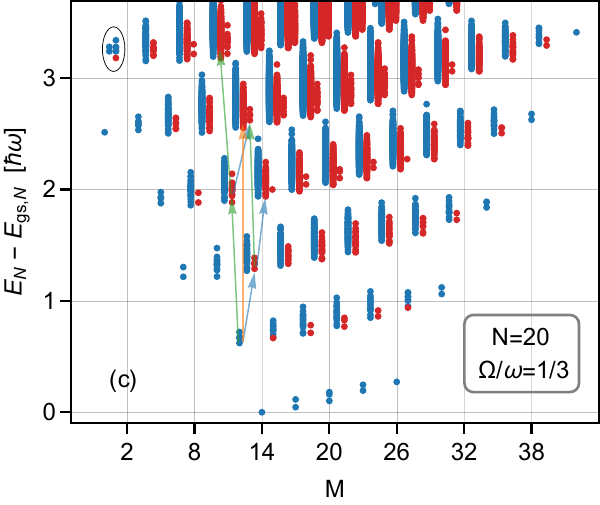}\label{fig:8c}}\qquad
\subfigure{\includegraphics[scale=0.8]{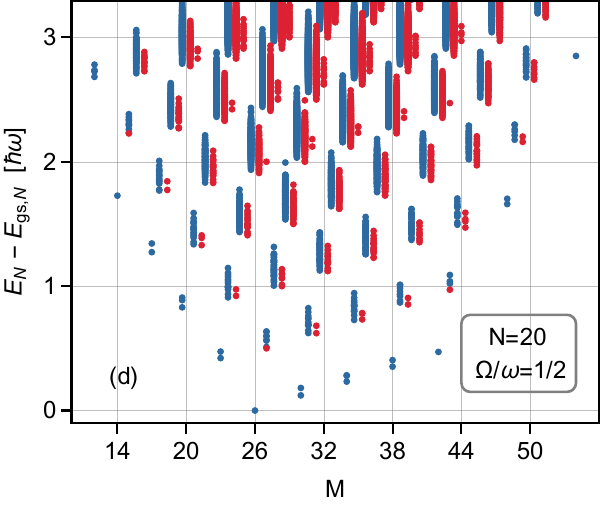}\label{fig:8d}}
\caption{
Excitation energies for $N=20$ particles in a rotating harmonic trap at rotation frequencies $\Omega/\omega=0,1/4,1/3$, and $1/2$, ordered by angular momentum for an attractive interaction \mbox{$g=-1$}. The color coding is the same as in Fig.~\ref{fig:7}: Primary states (nonprimary states) are represented by blue (red) points  and overlapping points are moved horizontally for clarity. In (c), we indicate by arrows the first few states of the conformal tower originating from an excited primary state at $M=12$  (compare with Fig.~\ref{fig:2}). The circled states in panels (a)--(c) show the evolution of a cluster of states under rotations.
}
\label{fig:8}
\end{figure*}

In this section, we explicitly confirm the conformal tower structure outlined in the previous section by exact diagonalization of the energy spectrum within degenerate first-order perturbation theory. To this end, we construct for a given particle number $N$ the ground and excited state manifolds with equal noninteracting energy~$E_N^{(0)}$ and diagonalize the Hamiltonian $H(\Omega)$ [Eq.~\eqref{eq:Ham}] as discussed in Sec.~\ref{sec:2}. In the diagonalization, we include the total spin operator $S^2$~[Eq.~\eqref{eq:spin}], the angular momentum operator $L_z$, and the Casimir operator $T$~[Eq.~\eqref{eq:casimir}],
\begin{equation} \label{eq:diagX}
W=\gamma_1 H(\Omega) + \gamma_2 L_z + \gamma_3 S^2 + \gamma_4 T 
\end{equation}
with incommensurate coefficients $\{\gamma_i\}$. 
Diagonalizing the matrix $W$ then gives simultaneous eigenstates of all (commuting) operators, and we determine the eigenvalues of the individual operators in Eq.~\eqref{eq:diagX} by computing their expectation values with the obtained eigenstates.
Primary states and their excited nonprimary states are identified by the first integers $(a,b,c)$ for which an eigenstate is in the kernel of the operators $R^{a+1}$, $Q_+^{b+1}$, and $Q_-^{c+1}$, which connect different degenerate subspaces. 

Note that care must be taken when applying this procedure to determine the indices $(a,b,c)$ of the nonprimary states, since
states $R^\dagger|P\rangle$ and $Q_+^\dagger Q_-^\dagger|P\rangle$ within the same conformal tower share eigenvalues of all operators (cf.~Fig.~\ref{fig:2}) (the same applies to higher breathing mode excitations). Hence, any linear combination $ \alpha R^\dagger|P\rangle + \beta Q_+^\dagger Q_-^\dagger|P\rangle$ is a valid eigenstate of~\eqref{eq:diagX}, independent of the coefficients $\{ \gamma_i \}$, and the kernel condition can overcount the indices ($a,b,c$) for the small subset of such states. To disentangle different nonprimary states, we thus successively apply the diagonalization procedure to degenerate manifolds with increasing excitation energy and store the primary states. Nonprimary states at higher excitation energy are then constructed in a different way by acting on a lower-level primary state with the operators $R^{\dagger}$, $Q_+^{\dagger}$, and $Q_-^{\dagger}$. In all cases, we were able to confirm that these states are identical to the nonprimary states obtained by explicit diagonalization of~\eqref{eq:diagX}, which provides a check and confirmation of our analysis and shows for the degenerate subspace with overcounted kernel states that they span the same vector subspace.

In Figs.~\ref{fig:7} and~\ref{fig:8}, we show the results for the excitation spectrum obtained from degenerate perturbation theory for particle numbers $N=6$~(Fig.~\ref{fig:7}) and $20$~(Fig.~\ref{fig:8}) at four rotation frequencies $\Omega/\omega=0$, 1/4, 1/3, and 1/2 [panels (a)--(d)] grouped by angular momentum eigenvalue~\mbox{$M=\langle L_z \rangle$}. As discussed, without interactions most states are highly degenerate but interactions lift this degeneracy and split the spectrum. We visualize the split spectrum using an attractive interaction strength \mbox{$g=-1$} such that states are still clustered around their noninteracting excitation energies. In the figures, blue points represent primary states and red points nonprimary states, and we do not indicate different conformal towers for clarity. In addition, we shift primary states to the left and nonprimary states to the right of their angular momentum eigenvalue, and we separate degenerate states horizontally. Note that while the distribution of nonprimary states is dictated by the nonrelativistic conformal symmetry (and confirmed in our numerics), the primary states and their energies are specific to the theory and determined from our numerical results. 
The figures show the excitation spectrum with respect to the ground state, which changes as the rotation frequency is increased (cf.~the discussion in Sec.~\ref{sec:IIA}): For both particle numbers \mbox{$N=6$} and \mbox{$N=20$}, the ground state changes from a zero angular momentum state \mbox{$M=0$} to a state with finite angular momentum as the rotation frequency is increased further. For \mbox{$N=6$}, the finite-angular momentum ground state configuration is of the type shown in Fig.~\ref{fig:4}(d), where all spins occupy the lowest angular momentum single-particle states with $n_j=0$. As a consequence, it will remain the ground state at faster rotations. For \mbox{$N=20$}, the change to an \mbox{$M=14$} ground state [Fig.~\ref{fig:8}(b)] corresponds to moving a pair of opposite spins from a single-particle state with $n_j=3$ to the lowest unoccupied angular momentum single-particle state with $n_j=0$, and to \mbox{$M=26$} [Fig.~\ref{fig:8}(d)] by moving a pair from $n_j=2$. As the rotation frequency is  increased further, subsequent ground states have angular momentum $M=42,54,70$, and $90$.

Comparing different panels in Figs.~\ref{fig:7} and~\ref{fig:8}, the same subclusters are seen in the energy spectrum at different rotation frequencies. Consider, for example, the cluster of states at \mbox{$M=\pm 1$} around excitation energy 1 in Fig.~\ref{fig:8a} (circled states): As the rotation frequency increases, (b) and (c), these clusters shift in energy but their relative energy is unchanged. The same feature is seen for all other clusters at a given $M$ that are split by the interactions: Since the interaction matrix elements~\eqref{eq:matrixelement} conserve total angular momentum, the corrections $E_N^{(1)}$ are independent of rotation frequency and the only change with rotation is an overall shift in the position of the clusters by $-(M-M_0) \Omega$, where $M_0$ is the total angular momentum of the reference ground state in the figure (which, as discussed above, changes with~$\Omega$). 
Note that, as discussed in Sec.~\ref{sec:IIB}, this implies that our calculations are also valid in the limit of fast rotations \mbox{$\Omega\to 1^-$}, where $N$-particle states and their  excitations are restricted to the lowest Landau level. Since in this limit the noninteracting energy of a state is proportional to its total angular momentum, \mbox{$E_N^{(0)} = (1-\Omega) M$}, degenerate perturbation theory is equivalent to an exact diagonalization in a disk geometry restricted to the lowest Landau level~\cite{hofmann23}. In this context, a recent work by~\textcite{palm20} identifies spinful quantum Hall states in the level spectrum of rapidly rotating few-fermion ensembles in the lowest Landau level. By the arguments above, such few-body fractional quantum Hall states at rapid rotations are already present at finite rotation frequency: For example, the \mbox{$N=6$}-particle primary ground state with \mbox{$M=6$} shown in Fig.~\ref{fig:7}(d) can be identified with a \mbox{$(1,1,0)$}-Halperin state~\cite{halperin83,tong16,palm20}. Likewise, ferromagnetic and skyrmion ground states for rapid rotations with repulsive interactions~\cite{palm20}  appear in Figs.~\ref{fig:7} and~\ref{fig:8} as excited states with higher angular momentum. A detailed description of these lowest Landau level states is an exiting prospect for future work.

In all our calculations, we verify the energy spectrum as predicted by the nonrelativistic conformal and Galilean symmetry. For illustration, we indicate by arrows the first few states of the lowest conformal tower in Figs.~\ref{fig:7}(a)--7(c), where the corresponding primary state is the \mbox{$N=6$} particle ground state, which has \mbox{$M=0$}. The conformal tower structure sketched in Fig.~\ref{fig:2} and proven in Sec.~\ref{sec:primary} is clearly apparent, with fixed angular-momentum conserving breathing-mode excitations that do not depend on the rotation frequency (vertical arrows in every panel), and center-of-mass excitations that change the angular momentum and that depend on the rotation frequency (diagonal right and left arrows). The analogous conformal tower emanating from the ground primary state is also visible in Fig.~\ref{fig:8}. Of course, the ground state is not the only primary state, and a plethora of additional primary states emerges in the excitation spectra. For example, in Fig.~\ref{fig:7}(d) we find $943$ primary states out of $3023$ total states up to this excitation energy, and in Fig.~\ref{fig:8d} there are $10445$ primary states out of 17464 total states shown. For illustration, we indicate the first states in the conformal tower of an excited primary state at \mbox{$M=12$} (lowest state in the cluster of primary states) in Fig.~\ref{fig:8c}. As discussed, the primary states are unique to the conformal theory and are thus specific to the trapped Fermi gas. The primary eigenvectors themselves are independent of the rotation frequency, but their energy shifts by an amount set by their angular momentum.

\section{Internal Hyperradial Wave Function}\label{sec:hyperradial}

\begin{figure*}[t!]
\subfigure{\includegraphics[scale=0.53]{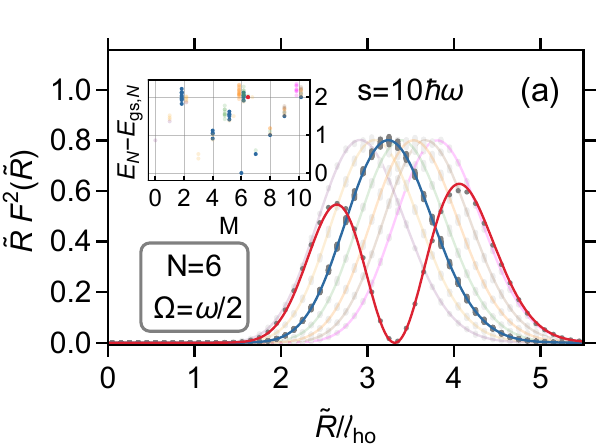}\label{fig:9a}} \qquad
\subfigure{\includegraphics[scale=0.53]{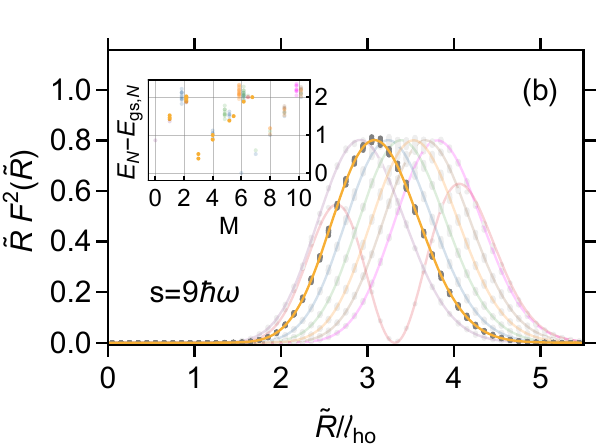}\label{fig:9b}} \qquad
\subfigure{\includegraphics[scale=0.53]{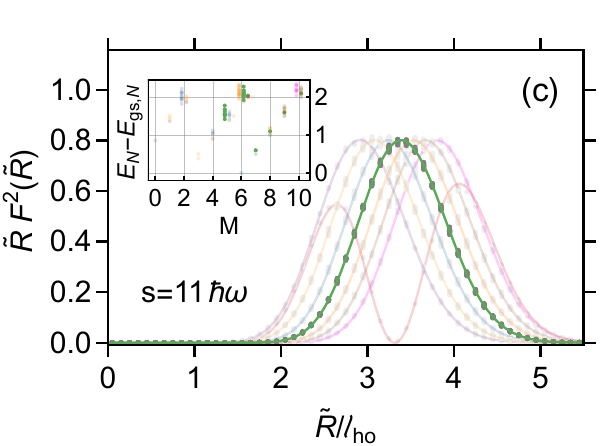}\label{fig:9c}} \\
\subfigure{\includegraphics[scale=0.53]{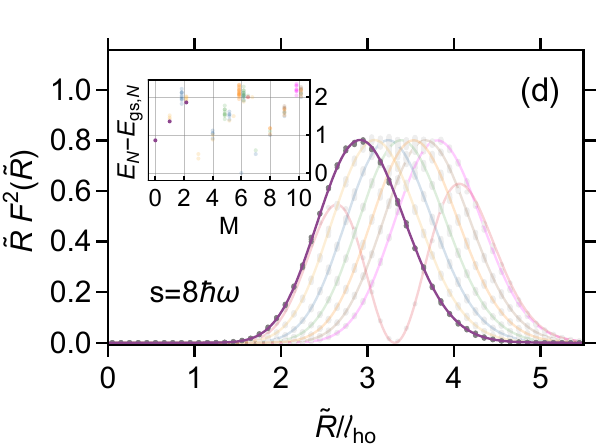}\label{fig:9d}} \qquad
\subfigure{\includegraphics[scale=0.53]{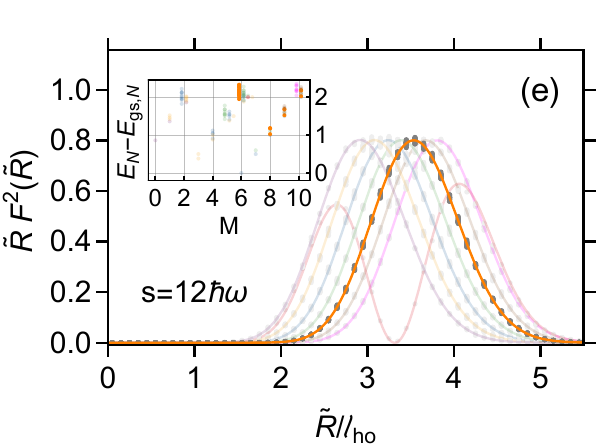}\label{fig:9e}} \qquad
\subfigure{\includegraphics[scale=0.53]{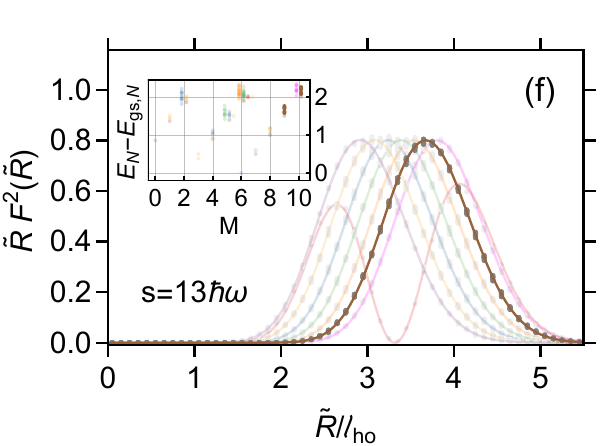}\label{fig:9f}} \\
\subfigure{\includegraphics[scale=0.53]{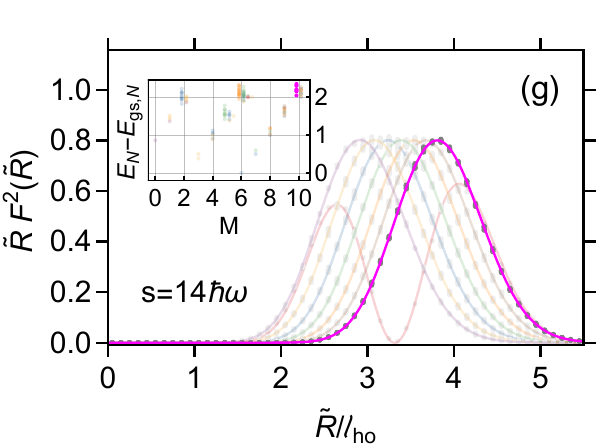}\label{fig:9g}}
\caption{Distribution of the internal hyperradius $\tilde{R}$, defined in Eq.~\eqref{eq:hypradius}, for the lowest 115 lowest eigenstates of \mbox{$N=6$} particles for $\Omega/\omega=1/2$. Gray points are the result of a Monte Carlo sampling of the many-body wave function, and continuous lines are the analytical prediction in Eq.~\eqref{eq:hyperradius}. Each figure highlights states with a particular value of the Casimir \mbox{$s=\sqrt{(\hbar\omega)^2+\langle T \rangle}$} for clarity. The insets show the same energy spectrum as in the highlighted region in Fig.~\ref{fig:7} but with a color coding that matches the hyperradial distribution.
}
\label{fig:9}
\end{figure*}

\begin{figure}[t!]
\subfigure{\includegraphics[scale=0.8]{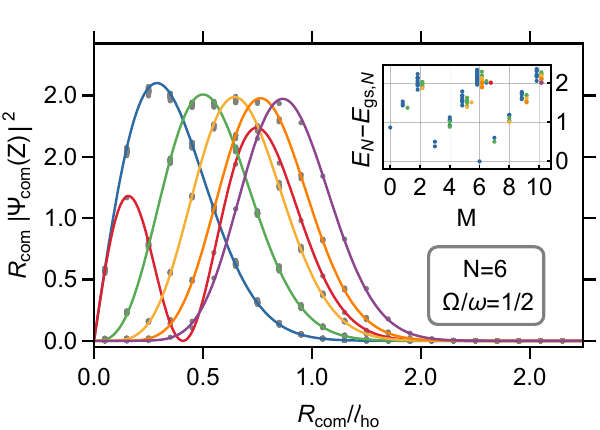}}
\caption{Distribution of the center-of-mass coordinate \mbox{$R_{\text{com}}=\sqrt{Z\bar{Z}}=\sqrt{(1/N\sum_{i\sigma} \mathbf{r}_{i\sigma} )^2}$} for the lowest 115 eigenstates of $N=6$ particles for $\Omega/\omega=1/2$. Gray points are the result of a Monte Carlo sampling of the many-body wave function, and continuous lines are the analytical prediction in Eq.~\eqref{eq:cm}. The inset shows the same energy spectrum as the highlighted region in Fig.~\ref{fig:7} but with a color coding matching the distribution.
}
\label{fig:10}
\end{figure}

On a microscopic level, the nonrelativistic conformal symmetry has its origin in a factorization of the many-body wave function~\cite{werner06,castin12}:
\begin{equation} \label{eq:wf}
\Psi({\bf r}_{1\uparrow}, \ldots, {\bf r}_{1\downarrow}, \ldots) = \Psi_{\text{com}}(Z) \,\frac{F(\tilde{R})}{\tilde{R}^{N-2}} \,\phi({\bf n}).
\end{equation}
Here, $\Psi_{\text{com}}(Z)$ is the center-of-mass part (which factorizes for any Galilean-invariant interaction), $F(\tilde{R})$ is the internal hyperradial part with hyperradius $\tilde{R}$ in Eq.~\eqref{eq:hypradius}, and $\phi({\bf n})$ is a hyperangular part that depends on the remaining internal coordinates \mbox{${\bf n} = (z_{1\uparrow} - Z, \ldots, z_{1\downarrow} - Z, \ldots)/\tilde{R}$}. We now confirm the hyperradial distribution using the eigenstates determined in the previous section.

The hyperradial distribution $F(\tilde{R})$ is  predicted by the conformal symmetry and determined for a state $|a,b,c\rangle_P$ by the condition \mbox{$(R)^{a+1}|a,b,c\rangle_P=0$}, yielding~\cite{bekassy22}
\begin{equation} 
F(\tilde{R}) = \sqrt{\frac{ 2 a!}{ \Gamma(s+a+1)}} \tilde{R}^s e^{-\tilde{R}^2/2} L_a^s(\tilde{R}^2) , \label{eq:hyperradius}
\end{equation}
where $\Gamma$ is the Gamma function (see Appendix~\ref{app:wf} for a derivation). The hyperradial wave function depends on the rank of the internal breathing mode excitation $a$, which sets the number of nodes in the wave function, but it does not depend on center-of-mass parameters $b$ and $c$ since these do not affect the internal dynamics. Furthermore, it depends on the Casimir parameter $s$ [Eq.~\eqref{eq:exp_cas}] that parametrizes the noninteracting energy of the corresponding primary state in the absence of rotations [Eq.~\eqref{eq:Eg_s}]. Thus, states that share the same expectation value of the Casimir operator have the same hyperradial distribution for a given number of breathing mode excitations. 

An experimentally observable consequence of the separability in Eq.~\eqref{eq:wf} is that $\tilde{R} F^2(\tilde{R})$ describes the distribution of the internal hyperradius~$\tilde{R}$, Eq.~\eqref{eq:hypradius}~\cite{blume07,castin12}. We confirm this using Metropolis Monte Carlo sampling of the perturbative wave function \mbox{$|\Psi_{a,b,c}({\bf r}_{1\uparrow}, \ldots, {\bf r}_{1\downarrow}, \ldots)|^2$} obtained from our diagonalization procedure. 
Details of the numerical implementation are described in Appendix~\ref{app:MonteC}. Figure~\ref{fig:9} shows results for the hyperradial distribution computed for the lowest 115 states of \mbox{$N=6$} particles with rotation frequency $\Omega=1/2$ [these states are highlighted in orange in Fig.~\ref{fig:7}(d)]. Gray points are Monte Carlo simulations and continuous lines are the analytical predictions in Eq.~\eqref{eq:hyperradius}, where the insets show the energy spectrum with a color coding matching the distributions. 
For clarity, the results for the hyperradial distribution are split into seven figures, where each figure highlights the results of a particular value of $s$, Eq.~\eqref{eq:Eg_s}, while the rest is opaque to allow a comparison between the figures. Since multiple states have equal Casimir, Eq.~\eqref{eq:exp_cas}, several distributions overlap. 
For example, Fig.~\ref{fig:9a} shows two distributions with equal value of \mbox{$s=10$}, where the blue curve is the distribution for 29 different states, and one red distribution corresponding to an internal breathing mode excitation of the ground state with $a=1$. Figures~\ref{fig:9b}--(g) show one distribution per plot, where all selected states are either primary states with equal $s$ or their center-of-mass excitations (cf. the figure insets): 16 states in Fig.~\ref{fig:9b}, 23 in Fig.~\ref{fig:9c}, 3 in Fig.~\ref{fig:9d}, 25 in Fig.~\ref{fig:9e}, 10 in Fig.~\ref{fig:9f}, and 8 in Fig.~\ref{fig:9g}. 
Note that the positions of the peaks in the distributions increase with increasing $s$, which parametrizes the energy of a primary state without rotation. Hence, in a rotating trap, the most compact distribution [Fig.~\ref{fig:9d}] corresponds to the ground state without rotations and its center-of-mass excitations  [i.e., states derived from the primary state in Fig.~\ref{fig:9d} with \mbox{$M=0$}], even though the former now forms an excited state in the rotating trap.

Different from the hyperradial distribution, the center-of-mass wave function $\Psi_{\text{com}}(Z)$ depends on $b$ and $c$, and is independent of internal dynamics and thus independent of the interaction potential. The wave function is determined by the relations \mbox{$(Q_+)^{b+1}|a,b,c\rangle_P=0 $} and \mbox{$(Q_-)^{c+1}|a,b,c\rangle_P=0 $}, yielding for $b\geq c$  (cf. Appendix~\ref{app:wf} for a derivation) 
\begin{align} \label{eq:cm}
\Psi_{\text {com}}(Z)=\sqrt{\frac{2 N^{1+b-c} c !}{b !}} Z^{b-c} e^{-N |Z|^2 / 2} L_c^{b-c}(N |Z|^2),
\end{align}
with $Z$ and $\bar{Z}$ as well as $b$ and $c$ exchanged for $c\geq b$. 
This is (up to normalization) exactly the wave function of a heavy particle with mass $Nm^*$ in an effective magnetic field \mbox{$B=2Nm^*\Omega$} in the $c$-th Landau level with angular momentum \mbox{$m=b-c$}~\cite{macdonald94,giuliani05}, again illustrating the interpretation of the center-of-mass modes as guiding-center and cyclotron modes.
The center-of-mass wave function~\eqref{eq:cm} depends on neither the energy of the primary state nor the number of breathing mode excitations $a$. 
Figure~\ref{fig:10} shows the center-of-mass distribution $R_{\text{com}} |\Psi_{\text{com}}(Z)|^2$ as a function of the modulus \mbox{$R_{\text{com}}=\sqrt{Z\bar{Z}}$}, where unlike in Fig.~\ref{fig:9} we avoid splitting the plots. As is apparent from the figure, states with the same $\text{max}(b,c)$ and $|b-c|$ share the same center-of-mass distribution. Hence, the distribution for all primary states as well as their internal breathing mode excitations, 64 states in total, collapse onto the blue curve. The red curve with a node corresponds to \mbox{$b=c=1$}, which describes one out of the 115 lowest states.

The hyperradial distribution (as well as the center-of-mass distribution) should be observable experimentally, hence verifying the conformal symmetry on a microscopic level, by sampling the many-body wave function with recently developed single-particle imaging techniques~\cite{holten21a,holten21b}. Deviations from our predictions are expected for stronger interactions, corresponding to anomalous symmetry breaking, or deformed traps, corresponding to introducing different length scales and explicitly breaking the symmetry.

\section{Conclusion} \label{sec:conclusion}

In this work, we have demonstrated that rotating 2D mesoscopic Fermi
gases at weak interactions possess a nonrelativistic conformal symmetry. We confirmed this by means of exact diagonalization of few-fermion ensembles in a harmonic trap, for which the conformal symmetry predicts so-called conformal towers formed by primary
states and their center-of-mass and internal breathing mode excitations, the latter having an
excitation energy at exactly twice the harmonic trap frequency. From the diagonalization, the eigenstates were used together with Monte Carlo simulations to compute and confirm the hyperradial distribution of the many-body wave function predicted by the symmetry. 
To the best of our knowledge, this provides the only setup, together with the nonrotating mesoscopic 2D Fermi gas considered in a previous work~\cite{bekassy22}, where the nonrelativistic conformal symmetry can be verified exactly by elementary means in an interacting quantum system. Thus, studying the rotating mesoscopic 2D Fermi gas can not only help our understanding of interacting systems in a magnetic field, but also give new insights into problems such as conformal nonequilibrium dynamics~\cite{bamler15,maki18,maki19,saintjalm19,lv20,shi20,olshanii21,maki2023emergent}.\\

{\it Note added:} Recently, the experimental work by~\textcite{lunt2024} appeared, which creates a two-particle Laughlin state in a rotating trap. This state corresponds to the lowest-lying \mbox{$N=2$} primary state with \mbox{$M=2$} discussed in this paper. The measurement of the two-body wave function using single-atom imagining is in excellent agreement with the prediction of this paper for the hyperradial wave function, Eq.~\eqref{eq:hyperradius}.

\begin{acknowledgments}
We thank Wilhelm Zwerger for discussions.  This work is supported by Vetenskapsr\aa det (Grant No. 2020-04239) (J.H.) and the Chalmers' Excellence Initiative Nano under its Excellence Ph.D. program (V.B.).
\end{acknowledgments}

\appendix

\section{DERIVATION OF THE CENTER-OF-MASS AND HYPERRADIAL WAVE FUNCTION} \label{app:wf}

In this Appendix we include an operator-based derivation of the center-of-mass wave function in Eq.~\eqref{eq:cm} and the hyperradial distribution in Eq.~\eqref{eq:hyperradius}. For the latter, the result is the same as in a nonrotating 3D trap, and an additional derivation can be found in the review by~\textcite{castin12}.

\subsection{Center-of-mass wave function}

The center-of-mass wave function in Eq.~\eqref{eq:cm} for an excited state $|a,b,c\rangle_P$ follows from the conditions
 \begin{align}
(Q_+)^{b+1}|a,b,c\rangle_P & =0 \label{app:eq:Qp} \\
(Q_-)^{c+1}|a,b,c\rangle_P & =0\label{app:eq:Qm} ,
\end{align}
which when written in a coordinate representation [Eqs.~\eqref{eq:defQ_Z1} and~\eqref{eq:defQ_Z2}] give two coupled differential equations. Denoting the center-of-mass wave function by \mbox{$\langle Z,\bar{Z} | a,b,c\rangle_P= \Psi_{b,c}(Z,\bar{Z})$}, we begin with the case \mbox{$b=c=0$}. The first condition~\eqref{app:eq:Qp} reads
\begin{align}
\langle Z,\bar{Z}| Q_+ |a,0,0 \rangle_P \sim \Bigl ( 2 \frac{\partial}{\partial Z} + N\bar{Z} \Bigr ) \Psi_{0,0}(Z,\bar{Z}) =0 ,
\end{align}
which implies $\Psi_{0,0}(Z,\bar{Z}) \sim f(\bar{Z})e^{-\frac{N|Z|^2}{2}}$, where $f(\bar{Z})$ is an arbitrary function of the $\bar{Z}$ coordinate. Likewise, inserting  $\Psi_{0,0}(Z,\bar{Z})$ in \eqref{app:eq:Qm} yields 
\begin{equation}
0=\langle Z,\bar{Z}| Q_- |a,0,0 \rangle_P \sim \Bigl ( 2 \frac{\partial}{\partial \bar{Z}} + NZ \Bigr ) f(\bar{Z})e^{-\frac{N|Z|^2}{2}},
\end{equation}
which implies $f(\bar{Z})=\text{const}$ and thus an (unnormalized) wave function \mbox{$ \Psi_{0,0}(Z,\bar{Z})=e^{-\frac{N|Z|^2}{2}}$}. 

The center-of-mass wave function of the excited state $|a,b,0 \rangle_P$ is obtained by acting with $(Q_+^{\dagger})^{b}$ on $\Psi_{0,0}(Z,\bar{Z})$:
\begin{align}
&\Psi_{b,0}(Z,\bar{Z}) = \langle Z,\bar{Z}| (Q_+^{\dagger })^b |a,0,0 \rangle_P \nonumber \\[1ex]
&\sim \Bigl(-2 \frac{\partial}{\partial \bar{Z}} + NZ\Bigr)^b e^{-\frac{N|Z|^2}{2}} \sim Z^b  e^{-\frac{N|Z|^2}{2}} .
\end{align}
Assuming $b\geq c$, we determine the general wave function by acting with $(Q_+^{\dagger})^{c}$ on $\Psi_{b,0}(Z,\bar{Z})$: 
\begin{align}
&\Psi_{b,c}(Z,\bar{Z}) = \langle Z,\bar{Z}| (Q_-^{\dagger })^c |a,b,0 \rangle \nonumber \\[1ex]
&\sim \Bigl(-2 \frac{\partial}{\partial Z} + N \bar{Z}\Bigr)^c  Z^b e^{-\frac{N|Z|^2}{2}} \nonumber \\
&\sim e^{-\frac{N|Z|^2}{2}} Z ^{b-c}\sum_{l=0}^c \frac{\left(\frac{-N|z|^2}{2}\right)^l}{l!} L_{c-l}^{(b-c)+l}\left(\frac{N|Z|^2}{2}\right) ,
\end{align}
where in the last line we expanded the prefactor using the binomial formula and used the Rodriguez representation of the associated Laguerre polynomials. 
Now using the recurrence relation for Laguerre polynomials, 
we  obtain the center-of-mass wave function (for \mbox{$b\geq c$}) stated in Eq.~\eqref{eq:cm} of the main text, 
which is normalized as \mbox{$\int d|Z| \, |Z| \, |\Psi_{\text {com}}(Z)|^2=1$}. For $c\geq b$, the derivation is analogous, but with $\bar{Z}$ replacing $Z$ as well as $c$ and $b$ exchanged.

\subsection{Hyperradial Distribution}

The hyperradial distribution in Eq.~\eqref{eq:hyperradius} for a state $|a,b,c\rangle_P$ is determined by the relation 
\begin{align}
R^{a+1}|a,b,c\rangle_P & =0\label{app:eq:R} .
\end{align}
It is useful to rewrite the $R^{\dagger}$ operator defined in Eq.~\eqref{eq:defR} as
\begin{equation}
R^{\dagger}=i D^{\text{int}} +H^{\text{int}} - 2 C^{\text{int}},
\end{equation}
with \mbox{$D^{\text{int}}=-i(N-1)-i \tilde{R}\partial_{\tilde{R}}$} denoting the generator of internal scale transformations, $H^{\text{int}}$ the internal Hamiltonian in the absence of rotation ($\Omega=0$), and $C^{\text{int}}=\tilde{R}^2/2$ the generator of internal special conformal transformations. Acting on a state $| a,b,c \rangle_P $ with $H^{\text{int}}$ yields the internal energy in a trap without rotation $s+1+2a$, where \mbox{$s=\sqrt{1+\langle T \rangle}$} and $T$ is the Casimir operator in Eq.~\eqref{eq:casimir}.

Defining
\begin{align}
\mathcal{F}_{a,s}(\tilde{R} ) = \frac{F_{a,s}(\tilde{R})}{\tilde{R}^{N-2}} = \langle \tilde{R} | a,b,c \rangle_P ,
\end{align}
the case $a=0$ follows from
\begin{equation}
\begin{split}
& \langle \tilde{R} | R |0,b,c \rangle_P \\
&= \Bigl ( -(N-1) -\tilde{R}\frac{\partial}{\partial \tilde{R}} +s+1 -\tilde{R}^2  \Bigr ) \mathcal{F}_{0,s}(\tilde{R} ) = 0 ,
\end{split}
\end{equation}
which gives the unnormalized hyperradial distribution \mbox{$F_{0,s}(\tilde{R})=\tilde{R}^s e^{-\tilde{R}^2/2} = \tilde{R}^s e^{-\tilde{R}^2/2} L_{0}^{s}(\tilde{R}^2)$}. The case $a=1$ is determined by acting with $R^{\dagger}$ on $|0,b,c\rangle_P$,
\begin{align}
\mathcal{F}_{a,s}(\tilde{R} ) &= \Bigl ( (N-1) +\tilde{R}\frac{\partial}{\partial \tilde{R}} +s+1 -\tilde{R}^2  \Bigr ) \frac{F_{0,s}(\tilde{R})}{\tilde{R}^{N-2}} \nonumber \\ 
&\sim \tilde{R}^s e^{\tilde{R}^2/2}\bigl( s+1-\tilde{R}^2\bigr)/ \tilde{R}^{N-2},
\end{align}
which implies~$F_{1,s}(\tilde{R})=\tilde{R}^s e^{\tilde{R}^2/2}( s+1-\tilde{R}^2) = \tilde{R}^s e^{\tilde{R}^2/2} L_{1}^{s}(\tilde{R}^2)$. 
Now, the unnormalized form for general~$a$, \mbox{$F_{a,s}(\tilde{R})= \tilde{R}^s e^{\tilde{R}^2/2} L_a^{s}(\tilde{R}^2)$}, follows by induction using the internal breathing mode excitation \mbox{$|a+1,b,c\rangle_P=R^{\dagger}|a,b,c\rangle_P$}. We have
\begin{align}
&\mathcal{F}_{a+1,s}(\tilde{R} ) \nonumber\\
&\sim \Bigl ( (N-1) +\tilde{R}\frac{\partial}{\partial \tilde{R}} +s+1+2a -\tilde{R}^2  \Bigr ) \mathcal{F}_{a,s}(\tilde{R} ) \nonumber \\
&= 2 \tilde{R}^{s-N+2}e^{-\frac{\tilde{R}^2}{2}} \Bigl[ (s+a+1-\tilde{R}^2)L_a^{s}(\tilde{R}^2) \nonumber\\
&\qquad\qquad\qquad -\tilde{R}^2L_{a-1}^{s+1}(\tilde{R}^2) \Bigr]. 
\end{align}
Using the recurrence relation for Laguerre polynomials, 
we end up with
\begin{equation}
F_{a+1,s}(\tilde{R}) \sim \tilde{R}^s e^{-\tilde{R}^2/2}L_{a+1}^{s}(\tilde{R}^2),
\end{equation}
as required. After normalizing $F_{a,s}(\tilde{R})$ according to $\int d \tilde{R} \, \tilde{R} \, F_{a,s}^2(\tilde{R}) = 1$, we obtain the hyperradial distribution in Eq.~\eqref{eq:hyperradius} of the main text.

\section{MONTE CARLO SAMPLING OF THE WAVE FUNCTIONS}\label{app:MonteC}

Having obtained the many-body wave function \mbox{$\Psi_{a,b,c}(\mathbf{r}_{1\uparrow},\ldots,\mathbf{r}_{1\downarrow},\ldots)$} from the diagonalization of \eqref{eq:Ham}, which is a superposition of Slater determinants, we employ Monte Carlo Metropolis sampling of the probability density \mbox{$|\Psi_{a,b,c}(\mathbf{r}_{1\uparrow},\ldots,\mathbf{r}_{1\downarrow},\ldots)|^2$} (see Ref. \cite{VBthesis}). The algorithm is initiated by randomly distributing the particle positions within a radius of $\ell_{ho}$ around the trap center. We then choose a particle at random and propose a new particle position, \mbox{$\mathbf{r}'_{j\sigma}=\mathbf{r}_{j\sigma}+\mathbf{r}$}, by a distance $r\in [0,\delta]$ with variance $\delta=0.7\ell_{ho}$, in a random direction, and 
compute the ratio
\begin{equation}
\xi=\frac{|\Psi_{a,b,c}(\ldots,\mathbf{r}_{j\sigma}',\ldots)|^2}{|\Psi_{a,b,c}(\ldots,\mathbf{r}_{j\sigma},\ldots)|^2}.
\end{equation}
We accept the new configuration if the ratio $\xi>\xi'$, where $\xi'\in [0,1]$ is a random number, otherwise we keep the initial configuration. For a given configuration, we compute the center-of-mass coordinate \mbox{$R_{\text{com}}=\sqrt{(1/N\sum_{i\sigma} \mathbf{r}_{i\sigma} )^2}$} and the hyperradius $\tilde{R}$ as defined in Eq.~\eqref{eq:hyperradius} to sample the distribution functions presented in Sec.~\ref{sec:hyperradial}. After a warm-up period of $10^4$ steps, every new proposed particle configuration is a sampling step, and we build histograms of $R_{\text{com}}$ and $\tilde{R}$ with bin size \mbox{$\Delta R_{\text{com}} = \Delta \tilde{R}=0.1\ell_{ho}$} for $10^6$ sampling steps. The results of this sampling procedure for selected $N=6$ particle wave functions are shown as gray points in Figs.~\ref{fig:9} and~\ref{fig:10}.

\bibliography{bib_rotating}

\begin{thebibliography}{69}%
\makeatletter
\providecommand \@ifxundefined [1]{%
 \@ifx{#1\undefined}
}%
\providecommand \@ifnum [1]{%
 \ifnum #1\expandafter \@firstoftwo
 \else \expandafter \@secondoftwo
 \fi
}%
\providecommand \@ifx [1]{%
 \ifx #1\expandafter \@firstoftwo
 \else \expandafter \@secondoftwo
 \fi
}%
\providecommand \natexlab [1]{#1}%
\providecommand \enquote  [1]{``#1''}%
\providecommand \bibnamefont  [1]{#1}%
\providecommand \bibfnamefont [1]{#1}%
\providecommand \citenamefont [1]{#1}%
\providecommand \href@noop [0]{\@secondoftwo}%
\providecommand \href [0]{\begingroup \@sanitize@url \@href}%
\providecommand \@href[1]{\@@startlink{#1}\@@href}%
\providecommand \@@href[1]{\endgroup#1\@@endlink}%
\providecommand \@sanitize@url [0]{\catcode `\\12\catcode `\$12\catcode
  `\&12\catcode `\#12\catcode `\^12\catcode `\_12\catcode `\%12\relax}%
\providecommand \@@startlink[1]{}%
\providecommand \@@endlink[0]{}%
\providecommand \url  [0]{\begingroup\@sanitize@url \@url }%
\providecommand \@url [1]{\endgroup\@href {#1}{\urlprefix }}%
\providecommand \urlprefix  [0]{URL }%
\providecommand \Eprint [0]{\href }%
\providecommand \doibase [0]{https://doi.org/}%
\providecommand \selectlanguage [0]{\@gobble}%
\providecommand \bibinfo  [0]{\@secondoftwo}%
\providecommand \bibfield  [0]{\@secondoftwo}%
\providecommand \translation [1]{[#1]}%
\providecommand \BibitemOpen [0]{}%
\providecommand \bibitemStop [0]{}%
\providecommand \bibitemNoStop [0]{.\EOS\space}%
\providecommand \EOS [0]{\spacefactor3000\relax}%
\providecommand \BibitemShut  [1]{\csname bibitem#1\endcsname}%
\let\auto@bib@innerbib\@empty
\bibitem [{\citenamefont {Madison}\ \emph {et~al.}(2000)\citenamefont
  {Madison}, \citenamefont {Chevy}, \citenamefont {Wohlleben},\ and\
  \citenamefont {Dalibard}}]{Madison00}%
  \BibitemOpen
  \bibfield  {author} {\bibinfo {author} {\bibfnamefont {K.~W.}\ \bibnamefont
  {Madison}}, \bibinfo {author} {\bibfnamefont {F.}~\bibnamefont {Chevy}},
  \bibinfo {author} {\bibfnamefont {W.}~\bibnamefont {Wohlleben}},\ and\
  \bibinfo {author} {\bibfnamefont {J.}~\bibnamefont {Dalibard}},\ }\bibfield
  {title} {\bibinfo {title} {{Vortex Formation in a Stirred Bose-Einstein
  Condensate}},\ }\href {https://doi.org/10.1103/PhysRevLett.84.806} {\bibfield
   {journal} {\bibinfo  {journal} {Phys. Rev. Lett.}\ }\textbf {\bibinfo
  {volume} {84}},\ \bibinfo {pages} {806} (\bibinfo {year} {2000})}\BibitemShut
  {NoStop}%
\bibitem [{\citenamefont {Abo-Shaeer}\ \emph {et~al.}(2001)\citenamefont
  {Abo-Shaeer}, \citenamefont {Raman}, \citenamefont {Vogels},\ and\
  \citenamefont {Ketterle}}]{aboshaer01}%
  \BibitemOpen
  \bibfield  {author} {\bibinfo {author} {\bibfnamefont {J.~R.}\ \bibnamefont
  {Abo-Shaeer}}, \bibinfo {author} {\bibfnamefont {C.}~\bibnamefont {Raman}},
  \bibinfo {author} {\bibfnamefont {J.~M.}\ \bibnamefont {Vogels}},\ and\
  \bibinfo {author} {\bibfnamefont {W.}~\bibnamefont {Ketterle}},\ }\bibfield
  {title} {\bibinfo {title} {{Observation of Vortex Lattices in Bose-Einstein
  Condensates}},\ }\href {https://doi.org/10.1126/science.1060182} {\bibfield
  {journal} {\bibinfo  {journal} {Science}\ }\textbf {\bibinfo {volume}
  {292}},\ \bibinfo {pages} {476} (\bibinfo {year} {2001})}\BibitemShut
  {NoStop}%
\bibitem [{\citenamefont {Zwierlein}\ \emph {et~al.}(2005)\citenamefont
  {Zwierlein}, \citenamefont {Abo-Shaeer}, \citenamefont {Schirotzek},
  \citenamefont {Schunck},\ and\ \citenamefont {Ketterle}}]{Zwierlein05}%
  \BibitemOpen
  \bibfield  {author} {\bibinfo {author} {\bibfnamefont {M.~W.}\ \bibnamefont
  {Zwierlein}}, \bibinfo {author} {\bibfnamefont {J.~R.}\ \bibnamefont
  {Abo-Shaeer}}, \bibinfo {author} {\bibfnamefont {A.}~\bibnamefont
  {Schirotzek}}, \bibinfo {author} {\bibfnamefont {C.~H.}\ \bibnamefont
  {Schunck}},\ and\ \bibinfo {author} {\bibfnamefont {W.}~\bibnamefont
  {Ketterle}},\ }\bibfield  {title} {\bibinfo {title} {{Vortices and
  superfluidity in a strongly interacting Fermi gas}},\ }\href
  {https://doi.org/10.1038/nature03858} {\bibfield  {journal} {\bibinfo
  {journal} {Nature}\ }\textbf {\bibinfo {volume} {435}},\ \bibinfo {pages}
  {1047} (\bibinfo {year} {2005})}\BibitemShut {NoStop}%
\bibitem [{\citenamefont {Fletcher}\ \emph {et~al.}(2021)\citenamefont
  {Fletcher}, \citenamefont {Shaffer}, \citenamefont {Wilson}, \citenamefont
  {Patel}, \citenamefont {Yan}, \citenamefont {Cr{\'e}pel}, \citenamefont
  {Mukherjee},\ and\ \citenamefont {Zwierlein}}]{Fletcher21}%
  \BibitemOpen
  \bibfield  {author} {\bibinfo {author} {\bibfnamefont {R.~J.}\ \bibnamefont
  {Fletcher}}, \bibinfo {author} {\bibfnamefont {A.}~\bibnamefont {Shaffer}},
  \bibinfo {author} {\bibfnamefont {C.~C.}\ \bibnamefont {Wilson}}, \bibinfo
  {author} {\bibfnamefont {P.~B.}\ \bibnamefont {Patel}}, \bibinfo {author}
  {\bibfnamefont {Z.}~\bibnamefont {Yan}}, \bibinfo {author} {\bibfnamefont
  {V.}~\bibnamefont {Cr{\'e}pel}}, \bibinfo {author} {\bibfnamefont
  {B.}~\bibnamefont {Mukherjee}},\ and\ \bibinfo {author} {\bibfnamefont
  {M.~W.}\ \bibnamefont {Zwierlein}},\ }\bibfield  {title} {\bibinfo {title}
  {Geometric squeezing into the lowest {Landau} level},\ }\href
  {https://doi.org/10.1126/science.aba7202} {\bibfield  {journal} {\bibinfo
  {journal} {Science}\ }\textbf {\bibinfo {volume} {372}},\ \bibinfo {pages}
  {1318} (\bibinfo {year} {2021})}\BibitemShut {NoStop}%
\bibitem [{\citenamefont {Cr\'epel}\ \emph {et~al.}(2023)\citenamefont
  {Cr\'epel}, \citenamefont {Yao}, \citenamefont {Mukherjee}, \citenamefont
  {Fletcher},\ and\ \citenamefont {Zwierlein}}]{Fletcher23}%
  \BibitemOpen
  \bibfield  {author} {\bibinfo {author} {\bibfnamefont {V.}~\bibnamefont
  {Cr\'epel}}, \bibinfo {author} {\bibfnamefont {R.}~\bibnamefont {Yao}},
  \bibinfo {author} {\bibfnamefont {B.}~\bibnamefont {Mukherjee}}, \bibinfo
  {author} {\bibfnamefont {R.}~\bibnamefont {Fletcher}},\ and\ \bibinfo
  {author} {\bibfnamefont {M.}~\bibnamefont {Zwierlein}},\ }\bibfield  {title}
  {\bibinfo {title} {Geometric squeezing of rotating quantum gases into the
  lowest {Landau} level},\ }\href {https://doi.org/10.5802/crphys.173}
  {\bibfield  {journal} {\bibinfo  {journal} {Comptes Rendus. Physique}\
  }\textbf {\bibinfo {volume} {24}},\ \bibinfo {pages} {241} (\bibinfo {year}
  {2023})}\BibitemShut {NoStop}%
\bibitem [{\citenamefont {Bloch}\ \emph {et~al.}(2008)\citenamefont {Bloch},
  \citenamefont {Dalibard},\ and\ \citenamefont {Zwerger}}]{bloch08}%
  \BibitemOpen
  \bibfield  {author} {\bibinfo {author} {\bibfnamefont {I.}~\bibnamefont
  {Bloch}}, \bibinfo {author} {\bibfnamefont {J.}~\bibnamefont {Dalibard}},\
  and\ \bibinfo {author} {\bibfnamefont {W.}~\bibnamefont {Zwerger}},\
  }\bibfield  {title} {\bibinfo {title} {{Many-body physics with ultracold
  gases}},\ }\href {https://doi.org/10.1103/RevModPhys.80.885} {\bibfield
  {journal} {\bibinfo  {journal} {Rev. Mod. Phys.}\ }\textbf {\bibinfo {volume}
  {80}},\ \bibinfo {pages} {885} (\bibinfo {year} {2008})}\BibitemShut
  {NoStop}%
\bibitem [{\citenamefont {Hofmann}\ and\ \citenamefont
  {Zwerger}(2023)}]{hofmann23}%
  \BibitemOpen
  \bibfield  {author} {\bibinfo {author} {\bibfnamefont {J.}~\bibnamefont
  {Hofmann}}\ and\ \bibinfo {author} {\bibfnamefont {W.}~\bibnamefont
  {Zwerger}},\ }\bibfield  {title} {\bibinfo {title} {Scale {Invariance} in the
  {Lowest} {Landau} {Level}},\ }\bibfield  {journal} {\bibinfo  {journal}
  {Comptes Rendus. Physique}\ }\href {https://doi.org/10.5802/crphys.137}
  {10.5802/crphys.137} (\bibinfo {year} {2023})\BibitemShut {NoStop}%
\bibitem [{\citenamefont {Olshanii}\ \emph {et~al.}(2010)\citenamefont
  {Olshanii}, \citenamefont {Perrin},\ and\ \citenamefont
  {Lorent}}]{olshanii10}%
  \BibitemOpen
  \bibfield  {author} {\bibinfo {author} {\bibfnamefont {M.}~\bibnamefont
  {Olshanii}}, \bibinfo {author} {\bibfnamefont {H.}~\bibnamefont {Perrin}},\
  and\ \bibinfo {author} {\bibfnamefont {V.}~\bibnamefont {Lorent}},\
  }\bibfield  {title} {\bibinfo {title} {{Example of a Quantum Anomaly in the
  Physics of Ultracold Gases}},\ }\href
  {https://doi.org/10.1103/PhysRevLett.105.095302} {\bibfield  {journal}
  {\bibinfo  {journal} {Phys. Rev. Lett.}\ }\textbf {\bibinfo {volume} {105}},\
  \bibinfo {pages} {095302} (\bibinfo {year} {2010})}\BibitemShut {NoStop}%
\bibitem [{\citenamefont {Vogt}\ \emph {et~al.}(2012)\citenamefont {Vogt},
  \citenamefont {Feld}, \citenamefont {Fr\"ohlich}, \citenamefont {Pertot},
  \citenamefont {Koschorreck},\ and\ \citenamefont {K\"ohl}}]{vogt12}%
  \BibitemOpen
  \bibfield  {author} {\bibinfo {author} {\bibfnamefont {E.}~\bibnamefont
  {Vogt}}, \bibinfo {author} {\bibfnamefont {M.}~\bibnamefont {Feld}}, \bibinfo
  {author} {\bibfnamefont {B.}~\bibnamefont {Fr\"ohlich}}, \bibinfo {author}
  {\bibfnamefont {D.}~\bibnamefont {Pertot}}, \bibinfo {author} {\bibfnamefont
  {M.}~\bibnamefont {Koschorreck}},\ and\ \bibinfo {author} {\bibfnamefont
  {M.}~\bibnamefont {K\"ohl}},\ }\bibfield  {title} {\bibinfo {title} {{Scale
  Invariance and Viscosity of a Two-Dimensional Fermi Gas}},\ }\href
  {https://doi.org/10.1103/PhysRevLett.108.070404} {\bibfield  {journal}
  {\bibinfo  {journal} {Phys. Rev. Lett.}\ }\textbf {\bibinfo {volume} {108}},\
  \bibinfo {pages} {070404} (\bibinfo {year} {2012})}\BibitemShut {NoStop}%
\bibitem [{\citenamefont {Gao}\ and\ \citenamefont {Yu}(2012)}]{gao12}%
  \BibitemOpen
  \bibfield  {author} {\bibinfo {author} {\bibfnamefont {C.}~\bibnamefont
  {Gao}}\ and\ \bibinfo {author} {\bibfnamefont {Z.}~\bibnamefont {Yu}},\
  }\bibfield  {title} {\bibinfo {title} {{Breathing mode of two-dimensional
  atomic Fermi gases in harmonic traps}},\ }\href
  {https://doi.org/10.1103/PhysRevA.86.043609} {\bibfield  {journal} {\bibinfo
  {journal} {Phys. Rev. A}\ }\textbf {\bibinfo {volume} {86}},\ \bibinfo
  {pages} {043609} (\bibinfo {year} {2012})}\BibitemShut {NoStop}%
\bibitem [{\citenamefont {Chafin}\ and\ \citenamefont
  {Sch\"afer}(2013)}]{chafin13}%
  \BibitemOpen
  \bibfield  {author} {\bibinfo {author} {\bibfnamefont {C.}~\bibnamefont
  {Chafin}}\ and\ \bibinfo {author} {\bibfnamefont {T.}~\bibnamefont
  {Sch\"afer}},\ }\bibfield  {title} {\bibinfo {title} {{Scale breaking and
  fluid dynamics in a dilute two-dimensional Fermi gas}},\ }\href
  {https://doi.org/10.1103/PhysRevA.88.043636} {\bibfield  {journal} {\bibinfo
  {journal} {Phys. Rev. A}\ }\textbf {\bibinfo {volume} {88}},\ \bibinfo
  {pages} {043636} (\bibinfo {year} {2013})}\BibitemShut {NoStop}%
\bibitem [{\citenamefont {Peppler}\ \emph {et~al.}(2018)\citenamefont
  {Peppler}, \citenamefont {Dyke}, \citenamefont {Zamorano}, \citenamefont
  {Herrera}, \citenamefont {Hoinka},\ and\ \citenamefont {Vale}}]{peppler18}%
  \BibitemOpen
  \bibfield  {author} {\bibinfo {author} {\bibfnamefont {T.}~\bibnamefont
  {Peppler}}, \bibinfo {author} {\bibfnamefont {P.}~\bibnamefont {Dyke}},
  \bibinfo {author} {\bibfnamefont {M.}~\bibnamefont {Zamorano}}, \bibinfo
  {author} {\bibfnamefont {I.}~\bibnamefont {Herrera}}, \bibinfo {author}
  {\bibfnamefont {S.}~\bibnamefont {Hoinka}},\ and\ \bibinfo {author}
  {\bibfnamefont {C.~J.}\ \bibnamefont {Vale}},\ }\bibfield  {title} {\bibinfo
  {title} {{Quantum Anomaly and 2D-3D Crossover in Strongly Interacting Fermi
  Gases}},\ }\href {https://doi.org/10.1103/PhysRevLett.121.120402} {\bibfield
  {journal} {\bibinfo  {journal} {Phys. Rev. Lett.}\ }\textbf {\bibinfo
  {volume} {121}},\ \bibinfo {pages} {120402} (\bibinfo {year}
  {2018})}\BibitemShut {NoStop}%
\bibitem [{\citenamefont {Holten}\ \emph {et~al.}(2018)\citenamefont {Holten},
  \citenamefont {Bayha}, \citenamefont {Klein}, \citenamefont {Murthy},
  \citenamefont {Preiss},\ and\ \citenamefont {Jochim}}]{holten18}%
  \BibitemOpen
  \bibfield  {author} {\bibinfo {author} {\bibfnamefont {M.}~\bibnamefont
  {Holten}}, \bibinfo {author} {\bibfnamefont {L.}~\bibnamefont {Bayha}},
  \bibinfo {author} {\bibfnamefont {A.~C.}\ \bibnamefont {Klein}}, \bibinfo
  {author} {\bibfnamefont {P.~A.}\ \bibnamefont {Murthy}}, \bibinfo {author}
  {\bibfnamefont {P.~M.}\ \bibnamefont {Preiss}},\ and\ \bibinfo {author}
  {\bibfnamefont {S.}~\bibnamefont {Jochim}},\ }\bibfield  {title} {\bibinfo
  {title} {{Anomalous Breaking of Scale Invariance in a Two-Dimensional Fermi
  Gas}},\ }\href {https://doi.org/10.1103/PhysRevLett.121.120401} {\bibfield
  {journal} {\bibinfo  {journal} {Phys. Rev. Lett.}\ }\textbf {\bibinfo
  {volume} {121}},\ \bibinfo {pages} {120401} (\bibinfo {year}
  {2018})}\BibitemShut {NoStop}%
\bibitem [{\citenamefont {Drut}\ \emph {et~al.}(2018)\citenamefont {Drut},
  \citenamefont {McKenney}, \citenamefont {Daza}, \citenamefont {Lin},\ and\
  \citenamefont {Ord\'o\~nez}}]{drut18}%
  \BibitemOpen
  \bibfield  {author} {\bibinfo {author} {\bibfnamefont {J.~E.}\ \bibnamefont
  {Drut}}, \bibinfo {author} {\bibfnamefont {J.~R.}\ \bibnamefont {McKenney}},
  \bibinfo {author} {\bibfnamefont {W.~S.}\ \bibnamefont {Daza}}, \bibinfo
  {author} {\bibfnamefont {C.~L.}\ \bibnamefont {Lin}},\ and\ \bibinfo {author}
  {\bibfnamefont {C.~R.}\ \bibnamefont {Ord\'o\~nez}},\ }\bibfield  {title}
  {\bibinfo {title} {{Quantum Anomaly and Thermodynamics of One-Dimensional
  Fermions with Three-Body Interactions}},\ }\href
  {https://doi.org/10.1103/PhysRevLett.120.243002} {\bibfield  {journal}
  {\bibinfo  {journal} {Phys. Rev. Lett.}\ }\textbf {\bibinfo {volume} {120}},\
  \bibinfo {pages} {243002} (\bibinfo {year} {2018})}\BibitemShut {NoStop}%
\bibitem [{\citenamefont {Mulkerin}\ \emph {et~al.}(2018)\citenamefont
  {Mulkerin}, \citenamefont {Liu},\ and\ \citenamefont {Hu}}]{mulkerin18}%
  \BibitemOpen
  \bibfield  {author} {\bibinfo {author} {\bibfnamefont {B.~C.}\ \bibnamefont
  {Mulkerin}}, \bibinfo {author} {\bibfnamefont {X.-J.}\ \bibnamefont {Liu}},\
  and\ \bibinfo {author} {\bibfnamefont {H.}~\bibnamefont {Hu}},\ }\bibfield
  {title} {\bibinfo {title} {{Collective modes of a two-dimensional Fermi gas
  at finite temperature}},\ }\href {https://doi.org/10.1103/PhysRevA.97.053612}
  {\bibfield  {journal} {\bibinfo  {journal} {Phys. Rev. A}\ }\textbf {\bibinfo
  {volume} {97}},\ \bibinfo {pages} {053612} (\bibinfo {year}
  {2018})}\BibitemShut {NoStop}%
\bibitem [{\citenamefont {Daza}\ \emph {et~al.}(2018)\citenamefont {Daza},
  \citenamefont {Drut}, \citenamefont {Lin},\ and\ \citenamefont
  {Ord\'o\~nez}}]{daza18}%
  \BibitemOpen
  \bibfield  {author} {\bibinfo {author} {\bibfnamefont {W.}~\bibnamefont
  {Daza}}, \bibinfo {author} {\bibfnamefont {J.~E.}\ \bibnamefont {Drut}},
  \bibinfo {author} {\bibfnamefont {C.}~\bibnamefont {Lin}},\ and\ \bibinfo
  {author} {\bibfnamefont {C.}~\bibnamefont {Ord\'o\~nez}},\ }\bibfield
  {title} {\bibinfo {title} {{Virial expansion for the Tan contact and
  Beth-Uhlenbeck formula from two-dimensional SO(2,1) anomalies}},\ }\href
  {https://doi.org/10.1103/PhysRevA.97.033630} {\bibfield  {journal} {\bibinfo
  {journal} {Phys. Rev. A}\ }\textbf {\bibinfo {volume} {97}},\ \bibinfo
  {pages} {033630} (\bibinfo {year} {2018})}\BibitemShut {NoStop}%
\bibitem [{\citenamefont {Hu}\ \emph {et~al.}(2019)\citenamefont {Hu},
  \citenamefont {Mulkerin}, \citenamefont {Toniolo}, \citenamefont {He},\ and\
  \citenamefont {Liu}}]{hu19}%
  \BibitemOpen
  \bibfield  {author} {\bibinfo {author} {\bibfnamefont {H.}~\bibnamefont
  {Hu}}, \bibinfo {author} {\bibfnamefont {B.~C.}\ \bibnamefont {Mulkerin}},
  \bibinfo {author} {\bibfnamefont {U.}~\bibnamefont {Toniolo}}, \bibinfo
  {author} {\bibfnamefont {L.}~\bibnamefont {He}},\ and\ \bibinfo {author}
  {\bibfnamefont {X.-J.}\ \bibnamefont {Liu}},\ }\bibfield  {title} {\bibinfo
  {title} {{Reduced Quantum Anomaly in a Quasi-Two-Dimensional Fermi
  Superfluid: Significance of the Confinement-Induced Effective Range of
  Interactions}},\ }\href {https://doi.org/10.1103/PhysRevLett.122.070401}
  {\bibfield  {journal} {\bibinfo  {journal} {Phys. Rev. Lett.}\ }\textbf
  {\bibinfo {volume} {122}},\ \bibinfo {pages} {070401} (\bibinfo {year}
  {2019})}\BibitemShut {NoStop}%
\bibitem [{\citenamefont {Yin}\ \emph {et~al.}(2020)\citenamefont {Yin},
  \citenamefont {Hu},\ and\ \citenamefont {Liu}}]{yin20}%
  \BibitemOpen
  \bibfield  {author} {\bibinfo {author} {\bibfnamefont {X.~Y.}\ \bibnamefont
  {Yin}}, \bibinfo {author} {\bibfnamefont {H.}~\bibnamefont {Hu}},\ and\
  \bibinfo {author} {\bibfnamefont {X.-J.}\ \bibnamefont {Liu}},\ }\bibfield
  {title} {\bibinfo {title} {{Few-Body Perspective of a Quantum Anomaly in
  Two-Dimensional Fermi Gases}},\ }\href
  {https://doi.org/10.1103/PhysRevLett.124.013401} {\bibfield  {journal}
  {\bibinfo  {journal} {Phys. Rev. Lett.}\ }\textbf {\bibinfo {volume} {124}},\
  \bibinfo {pages} {013401} (\bibinfo {year} {2020})}\BibitemShut {NoStop}%
\bibitem [{\citenamefont {Hofmann}(2012)}]{hofmann12}%
  \BibitemOpen
  \bibfield  {author} {\bibinfo {author} {\bibfnamefont {J.}~\bibnamefont
  {Hofmann}},\ }\bibfield  {title} {\bibinfo {title} {{Quantum Anomaly,
  Universal Relations, and Breathing Mode of a Two-Dimensional Fermi Gas}},\
  }\href {https://doi.org/10.1103/PhysRevLett.108.185303} {\bibfield  {journal}
  {\bibinfo  {journal} {Phys. Rev. Lett.}\ }\textbf {\bibinfo {volume} {108}},\
  \bibinfo {pages} {185303} (\bibinfo {year} {2012})}\BibitemShut {NoStop}%
\bibitem [{\citenamefont {Langmack}\ \emph {et~al.}(2012)\citenamefont
  {Langmack}, \citenamefont {Barth}, \citenamefont {Zwerger},\ and\
  \citenamefont {Braaten}}]{langmack12}%
  \BibitemOpen
  \bibfield  {author} {\bibinfo {author} {\bibfnamefont {C.}~\bibnamefont
  {Langmack}}, \bibinfo {author} {\bibfnamefont {M.}~\bibnamefont {Barth}},
  \bibinfo {author} {\bibfnamefont {W.}~\bibnamefont {Zwerger}},\ and\ \bibinfo
  {author} {\bibfnamefont {E.}~\bibnamefont {Braaten}},\ }\bibfield  {title}
  {\bibinfo {title} {{Clock Shift in a Strongly Interacting Two-Dimensional
  Fermi Gas}},\ }\href {https://doi.org/10.1103/PhysRevLett.108.060402}
  {\bibfield  {journal} {\bibinfo  {journal} {Phys. Rev. Lett.}\ }\textbf
  {\bibinfo {volume} {108}},\ \bibinfo {pages} {060402} (\bibinfo {year}
  {2012})}\BibitemShut {NoStop}%
\bibitem [{\citenamefont {Son}(2007)}]{son07}%
  \BibitemOpen
  \bibfield  {author} {\bibinfo {author} {\bibfnamefont {D.~T.}\ \bibnamefont
  {Son}},\ }\bibfield  {title} {\bibinfo {title} {{Vanishing Bulk Viscosities
  and Conformal Invariance of the Unitary Fermi Gas}},\ }\href
  {https://doi.org/10.1103/PhysRevLett.98.020604} {\bibfield  {journal}
  {\bibinfo  {journal} {Phys. Rev. Lett.}\ }\textbf {\bibinfo {volume} {98}},\
  \bibinfo {pages} {020604} (\bibinfo {year} {2007})}\BibitemShut {NoStop}%
\bibitem [{\citenamefont {Hofmann}(2020)}]{hofmann20}%
  \BibitemOpen
  \bibfield  {author} {\bibinfo {author} {\bibfnamefont {J.}~\bibnamefont
  {Hofmann}},\ }\bibfield  {title} {\bibinfo {title} {{High-temperature
  expansion of the viscosity in interacting quantum gases}},\ }\href
  {https://doi.org/10.1103/PhysRevA.101.013620} {\bibfield  {journal} {\bibinfo
   {journal} {Phys. Rev. A}\ }\textbf {\bibinfo {volume} {101}},\ \bibinfo
  {pages} {013620} (\bibinfo {year} {2020})}\BibitemShut {NoStop}%
\bibitem [{\citenamefont {Enss}(2019)}]{enss19}%
  \BibitemOpen
  \bibfield  {author} {\bibinfo {author} {\bibfnamefont {T.}~\bibnamefont
  {Enss}},\ }\bibfield  {title} {\bibinfo {title} {{Bulk Viscosity and Contact
  Correlations in Attractive Fermi Gases}},\ }\href
  {https://doi.org/10.1103/PhysRevLett.123.205301} {\bibfield  {journal}
  {\bibinfo  {journal} {Phys. Rev. Lett.}\ }\textbf {\bibinfo {volume} {123}},\
  \bibinfo {pages} {205301} (\bibinfo {year} {2019})}\BibitemShut {NoStop}%
\bibitem [{\citenamefont {Nishida}(2019)}]{nishida19}%
  \BibitemOpen
  \bibfield  {author} {\bibinfo {author} {\bibfnamefont {Y.}~\bibnamefont
  {Nishida}},\ }\bibfield  {title} {\bibinfo {title} {{Viscosity spectral
  functions of resonating fermions in the quantum virial expansion}},\ }\href
  {https://doi.org/https://doi.org/10.1016/j.aop.2019.167949} {\bibfield
  {journal} {\bibinfo  {journal} {Annals of Physics}\ }\textbf {\bibinfo
  {volume} {410}},\ \bibinfo {pages} {167949} (\bibinfo {year}
  {2019})}\BibitemShut {NoStop}%
\bibitem [{\citenamefont {Bekassy}\ and\ \citenamefont
  {Hofmann}(2022)}]{bekassy22}%
  \BibitemOpen
  \bibfield  {author} {\bibinfo {author} {\bibfnamefont {V.}~\bibnamefont
  {Bekassy}}\ and\ \bibinfo {author} {\bibfnamefont {J.}~\bibnamefont
  {Hofmann}},\ }\bibfield  {title} {\bibinfo {title} {Nonrelativistic conformal
  invariance in mesoscopic two-dimensional {F}ermi gases},\ }\href
  {https://doi.org/10.1103/PhysRevLett.128.193401} {\bibfield  {journal}
  {\bibinfo  {journal} {Phys. Rev. Lett.}\ }\textbf {\bibinfo {volume} {128}},\
  \bibinfo {pages} {193401} (\bibinfo {year} {2022})}\BibitemShut {NoStop}%
\bibitem [{\citenamefont {Hagen}(1972)}]{Hagen72}%
  \BibitemOpen
  \bibfield  {author} {\bibinfo {author} {\bibfnamefont {C.~R.}\ \bibnamefont
  {Hagen}},\ }\bibfield  {title} {\bibinfo {title} {{Scale and Conformal
  Transformations in Galilean-Covariant Field Theory}},\ }\href
  {https://doi.org/10.1103/PhysRevD.5.377} {\bibfield  {journal} {\bibinfo
  {journal} {Phys. Rev. D}\ }\textbf {\bibinfo {volume} {5}},\ \bibinfo {pages}
  {377} (\bibinfo {year} {1972})}\BibitemShut {NoStop}%
\bibitem [{\citenamefont {Zwerger}(2021)}]{zwerger21}%
  \BibitemOpen
  \bibfield  {author} {\bibinfo {author} {\bibfnamefont {W.}~\bibnamefont
  {Zwerger}},\ }\bibfield  {title} {\bibinfo {title} {{Basic Concepts and some
  current Directions in Ultracold Gases}},\ }\href@noop {} {\bibfield
  {journal} {\bibinfo  {journal} {{Lectures on many-body phenomena in ultracold
  gases, Coll{\`e}ge de France}}\ } (\bibinfo {year} {2021})}\BibitemShut
  {NoStop}%
\bibitem [{\citenamefont {Pitaevskii}\ and\ \citenamefont
  {Rosch}(1997)}]{pitaevskii97}%
  \BibitemOpen
  \bibfield  {author} {\bibinfo {author} {\bibfnamefont {L.~P.}\ \bibnamefont
  {Pitaevskii}}\ and\ \bibinfo {author} {\bibfnamefont {A.}~\bibnamefont
  {Rosch}},\ }\bibfield  {title} {\bibinfo {title} {{Breathing modes and hidden
  symmetry of trapped atoms in two dimensions}},\ }\href
  {https://doi.org/10.1103/PhysRevA.55.R853} {\bibfield  {journal} {\bibinfo
  {journal} {Phys. Rev. A}\ }\textbf {\bibinfo {volume} {55}},\ \bibinfo
  {pages} {R853} (\bibinfo {year} {1997})}\BibitemShut {NoStop}%
\bibitem [{\citenamefont {Castin}(2004)}]{castin04}%
  \BibitemOpen
  \bibfield  {author} {\bibinfo {author} {\bibfnamefont {Y.}~\bibnamefont
  {Castin}},\ }\bibfield  {title} {\bibinfo {title} {{Exact scaling transform
  for a unitary quantum gas in a time dependent harmonic potential}},\ }\href
  {https://doi.org/https://doi.org/10.1016/j.crhy.2004.03.017} {\bibfield
  {journal} {\bibinfo  {journal} {Comptes Rendus Physique}\ }\textbf {\bibinfo
  {volume} {5}},\ \bibinfo {pages} {407} (\bibinfo {year} {2004})}\BibitemShut
  {NoStop}%
\bibitem [{\citenamefont {Werner}\ and\ \citenamefont
  {Castin}(2006)}]{werner06}%
  \BibitemOpen
  \bibfield  {author} {\bibinfo {author} {\bibfnamefont {F.}~\bibnamefont
  {Werner}}\ and\ \bibinfo {author} {\bibfnamefont {Y.}~\bibnamefont
  {Castin}},\ }\bibfield  {title} {\bibinfo {title} {{Unitary gas in an
  isotropic harmonic trap: Symmetry properties and applications}},\ }\href
  {https://doi.org/10.1103/PhysRevA.74.053604} {\bibfield  {journal} {\bibinfo
  {journal} {Phys. Rev. A}\ }\textbf {\bibinfo {volume} {74}},\ \bibinfo
  {pages} {053604} (\bibinfo {year} {2006})}\BibitemShut {NoStop}%
\bibitem [{\citenamefont {Castin}\ and\ \citenamefont
  {Werner}(2012)}]{castin12}%
  \BibitemOpen
  \bibfield  {author} {\bibinfo {author} {\bibfnamefont {Y.}~\bibnamefont
  {Castin}}\ and\ \bibinfo {author} {\bibfnamefont {F.}~\bibnamefont
  {Werner}},\ }\bibfield  {title} {\bibinfo {title} {{The Unitary Gas and its
  Symmetry Properties}},\ }in\ \href
  {https://doi.org/10.1007/978-3-642-21978-8_5} {\emph {\bibinfo {booktitle}
  {{The BCS--BEC Crossover and the Unitary Fermi Gas}}}},\ \bibinfo {editor}
  {edited by\ \bibinfo {editor} {\bibfnamefont {W.}~\bibnamefont {Zwerger}}}\
  (\bibinfo  {publisher} {Springer (Heidelberg)},\ \bibinfo {year}
  {2012})\BibitemShut {NoStop}%
\bibitem [{\citenamefont {Mehen}\ \emph {et~al.}(2000)\citenamefont {Mehen},
  \citenamefont {Stewart},\ and\ \citenamefont {Wise}}]{mehen00}%
  \BibitemOpen
  \bibfield  {author} {\bibinfo {author} {\bibfnamefont {T.}~\bibnamefont
  {Mehen}}, \bibinfo {author} {\bibfnamefont {I.~W.}\ \bibnamefont {Stewart}},\
  and\ \bibinfo {author} {\bibfnamefont {M.~B.}\ \bibnamefont {Wise}},\
  }\bibfield  {title} {\bibinfo {title} {{Conformal invariance for
  non-relativistic field theory}},\ }\href
  {https://doi.org/https://doi.org/10.1016/S0370-2693(00)00006-X} {\bibfield
  {journal} {\bibinfo  {journal} {Physics Letters B}\ }\textbf {\bibinfo
  {volume} {474}},\ \bibinfo {pages} {145} (\bibinfo {year}
  {2000})}\BibitemShut {NoStop}%
\bibitem [{\citenamefont {Son}\ and\ \citenamefont {Wingate}(2006)}]{son06}%
  \BibitemOpen
  \bibfield  {author} {\bibinfo {author} {\bibfnamefont {D.~T.}\ \bibnamefont
  {Son}}\ and\ \bibinfo {author} {\bibfnamefont {M.}~\bibnamefont {Wingate}},\
  }\bibfield  {title} {\bibinfo {title} {{General coordinate invariance and
  conformal invariance in nonrelativistic physics: Unitary Fermi gas}},\ }\href
  {https://doi.org/https://doi.org/10.1016/j.aop.2005.11.001} {\bibfield
  {journal} {\bibinfo  {journal} {Annals of Physics}\ }\textbf {\bibinfo
  {volume} {321}},\ \bibinfo {pages} {197} (\bibinfo {year}
  {2006})}\BibitemShut {NoStop}%
\bibitem [{\citenamefont {Nishida}\ and\ \citenamefont
  {Son}(2007)}]{nishida07}%
  \BibitemOpen
  \bibfield  {author} {\bibinfo {author} {\bibfnamefont {Y.}~\bibnamefont
  {Nishida}}\ and\ \bibinfo {author} {\bibfnamefont {D.~T.}\ \bibnamefont
  {Son}},\ }\bibfield  {title} {\bibinfo {title} {{Nonrelativistic conformal
  field theories}},\ }\href {https://doi.org/10.1103/PhysRevD.76.086004}
  {\bibfield  {journal} {\bibinfo  {journal} {Phys. Rev. D}\ }\textbf {\bibinfo
  {volume} {76}},\ \bibinfo {pages} {086004} (\bibinfo {year}
  {2007})}\BibitemShut {NoStop}%
\bibitem [{\citenamefont {Bergschneider}\ \emph {et~al.}(2018)\citenamefont
  {Bergschneider}, \citenamefont {Klinkhamer}, \citenamefont {Becher},
  \citenamefont {Klemt}, \citenamefont {Z\"urn}, \citenamefont {Preiss},\ and\
  \citenamefont {Jochim}}]{Bergschneider18}%
  \BibitemOpen
  \bibfield  {author} {\bibinfo {author} {\bibfnamefont {A.}~\bibnamefont
  {Bergschneider}}, \bibinfo {author} {\bibfnamefont {V.~M.}\ \bibnamefont
  {Klinkhamer}}, \bibinfo {author} {\bibfnamefont {J.~H.}\ \bibnamefont
  {Becher}}, \bibinfo {author} {\bibfnamefont {R.}~\bibnamefont {Klemt}},
  \bibinfo {author} {\bibfnamefont {G.}~\bibnamefont {Z\"urn}}, \bibinfo
  {author} {\bibfnamefont {P.~M.}\ \bibnamefont {Preiss}},\ and\ \bibinfo
  {author} {\bibfnamefont {S.}~\bibnamefont {Jochim}},\ }\bibfield  {title}
  {\bibinfo {title} {Spin-resolved single-atom imaging of $^{6}\mathrm{Li}$ in
  free space},\ }\href {https://doi.org/10.1103/PhysRevA.97.063613} {\bibfield
  {journal} {\bibinfo  {journal} {Phys. Rev. A}\ }\textbf {\bibinfo {volume}
  {97}},\ \bibinfo {pages} {063613} (\bibinfo {year} {2018})}\BibitemShut
  {NoStop}%
\bibitem [{\citenamefont {Bayha}\ \emph {et~al.}(2020)\citenamefont {Bayha},
  \citenamefont {Holten}, \citenamefont {Klemt}, \citenamefont {Subramanian},
  \citenamefont {Bjerlin}, \citenamefont {Reimann}, \citenamefont {Bruun},
  \citenamefont {Preiss},\ and\ \citenamefont {Jochim}}]{bayha20}%
  \BibitemOpen
  \bibfield  {author} {\bibinfo {author} {\bibfnamefont {L.}~\bibnamefont
  {Bayha}}, \bibinfo {author} {\bibfnamefont {M.}~\bibnamefont {Holten}},
  \bibinfo {author} {\bibfnamefont {R.}~\bibnamefont {Klemt}}, \bibinfo
  {author} {\bibfnamefont {K.}~\bibnamefont {Subramanian}}, \bibinfo {author}
  {\bibfnamefont {J.}~\bibnamefont {Bjerlin}}, \bibinfo {author} {\bibfnamefont
  {S.~M.}\ \bibnamefont {Reimann}}, \bibinfo {author} {\bibfnamefont {G.~M.}\
  \bibnamefont {Bruun}}, \bibinfo {author} {\bibfnamefont {P.~M.}\ \bibnamefont
  {Preiss}},\ and\ \bibinfo {author} {\bibfnamefont {S.}~\bibnamefont
  {Jochim}},\ }\bibfield  {title} {\bibinfo {title} {{Observing the emergence
  of a quantum phase transition shell by shell}},\ }\href
  {https://doi.org/10.1038/s41586-020-2936-y} {\bibfield  {journal} {\bibinfo
  {journal} {Nature}\ }\textbf {\bibinfo {volume} {587}},\ \bibinfo {pages}
  {583} (\bibinfo {year} {2020})}\BibitemShut {NoStop}%
\bibitem [{\citenamefont {Holten}\ \emph {et~al.}(2021)\citenamefont {Holten},
  \citenamefont {Bayha}, \citenamefont {Subramanian}, \citenamefont {Heintze},
  \citenamefont {Preiss},\ and\ \citenamefont {Jochim}}]{holten21a}%
  \BibitemOpen
  \bibfield  {author} {\bibinfo {author} {\bibfnamefont {M.}~\bibnamefont
  {Holten}}, \bibinfo {author} {\bibfnamefont {L.}~\bibnamefont {Bayha}},
  \bibinfo {author} {\bibfnamefont {K.}~\bibnamefont {Subramanian}}, \bibinfo
  {author} {\bibfnamefont {C.}~\bibnamefont {Heintze}}, \bibinfo {author}
  {\bibfnamefont {P.~M.}\ \bibnamefont {Preiss}},\ and\ \bibinfo {author}
  {\bibfnamefont {S.}~\bibnamefont {Jochim}},\ }\bibfield  {title} {\bibinfo
  {title} {{Observation of Pauli Crystals}},\ }\href
  {https://doi.org/10.1103/PhysRevLett.126.020401} {\bibfield  {journal}
  {\bibinfo  {journal} {Phys. Rev. Lett.}\ }\textbf {\bibinfo {volume} {126}},\
  \bibinfo {pages} {020401} (\bibinfo {year} {2021})}\BibitemShut {NoStop}%
\bibitem [{\citenamefont {Holten}\ \emph {et~al.}(2022)\citenamefont {Holten},
  \citenamefont {Bayha}, \citenamefont {Subramanian}, \citenamefont
  {Brandstetter}, \citenamefont {Heintze}, \citenamefont {Lunt}, \citenamefont
  {Preiss},\ and\ \citenamefont {Jochim}}]{holten21b}%
  \BibitemOpen
  \bibfield  {author} {\bibinfo {author} {\bibfnamefont {M.}~\bibnamefont
  {Holten}}, \bibinfo {author} {\bibfnamefont {L.}~\bibnamefont {Bayha}},
  \bibinfo {author} {\bibfnamefont {K.}~\bibnamefont {Subramanian}}, \bibinfo
  {author} {\bibfnamefont {S.}~\bibnamefont {Brandstetter}}, \bibinfo {author}
  {\bibfnamefont {C.}~\bibnamefont {Heintze}}, \bibinfo {author} {\bibfnamefont
  {P.}~\bibnamefont {Lunt}}, \bibinfo {author} {\bibfnamefont {P.~M.}\
  \bibnamefont {Preiss}},\ and\ \bibinfo {author} {\bibfnamefont
  {S.}~\bibnamefont {Jochim}},\ }\bibfield  {title} {\bibinfo {title}
  {{Observation of Cooper pairs in a mesoscopic two-dimensional Fermi gas}},\
  }\href {https://doi.org/10.1038/s41586-022-04678-1} {\bibfield  {journal}
  {\bibinfo  {journal} {Nature}\ }\textbf {\bibinfo {volume} {606}},\ \bibinfo
  {pages} {287} (\bibinfo {year} {2022})}\BibitemShut {NoStop}%
\bibitem [{\citenamefont {Bekassy}(2023)}]{VBthesis}%
  \BibitemOpen
  \bibfield  {author} {\bibinfo {author} {\bibfnamefont {V.}~\bibnamefont
  {Bekassy}},\ }\emph {\bibinfo {title} {Rotating Two-Dimensional Mesoscopic
  {F}ermi Gases and Nonrelativistic Conformal Invariance}},\ \href
  {https://doi.org/20.500.12380/306203} {Master's thesis},\ \bibinfo  {school}
  {Chalmers University of Technology} (\bibinfo {year} {2023})\BibitemShut
  {NoStop}%
\bibitem [{\citenamefont {Slater}(1929)}]{slater1929}%
  \BibitemOpen
  \bibfield  {author} {\bibinfo {author} {\bibfnamefont {J.~C.}\ \bibnamefont
  {Slater}},\ }\bibfield  {title} {\bibinfo {title} {The theory of complex
  spectra},\ }\href {https://doi.org/10.1103/PhysRev.34.1293} {\bibfield
  {journal} {\bibinfo  {journal} {Phys. Rev.}\ }\textbf {\bibinfo {volume}
  {34}},\ \bibinfo {pages} {1293} (\bibinfo {year} {1929})}\BibitemShut
  {NoStop}%
\bibitem [{\citenamefont {Slater}(1931)}]{slater1931}%
  \BibitemOpen
  \bibfield  {author} {\bibinfo {author} {\bibfnamefont {J.~C.}\ \bibnamefont
  {Slater}},\ }\bibfield  {title} {\bibinfo {title} {Molecular energy levels
  and valence bonds},\ }\href {https://doi.org/10.1103/PhysRev.38.1109}
  {\bibfield  {journal} {\bibinfo  {journal} {Phys. Rev.}\ }\textbf {\bibinfo
  {volume} {38}},\ \bibinfo {pages} {1109} (\bibinfo {year}
  {1931})}\BibitemShut {NoStop}%
\bibitem [{\citenamefont {Condon}(1930)}]{condon1930}%
  \BibitemOpen
  \bibfield  {author} {\bibinfo {author} {\bibfnamefont {E.~U.}\ \bibnamefont
  {Condon}},\ }\bibfield  {title} {\bibinfo {title} {The theory of complex
  spectra},\ }\href {https://doi.org/10.1103/PhysRev.36.1121} {\bibfield
  {journal} {\bibinfo  {journal} {Phys. Rev.}\ }\textbf {\bibinfo {volume}
  {36}},\ \bibinfo {pages} {1121} (\bibinfo {year} {1930})}\BibitemShut
  {NoStop}%
\bibitem [{\citenamefont {L\"owdin}(1955)}]{lowdin55}%
  \BibitemOpen
  \bibfield  {author} {\bibinfo {author} {\bibfnamefont {P.-O.}\ \bibnamefont
  {L\"owdin}},\ }\bibfield  {title} {\bibinfo {title} {{Quantum Theory of
  Many-Particle Systems. I. Physical Interpretations by Means of Density
  Matrices, Natural Spin-Orbitals, and Convergence Problems in the Method of
  Configurational Interaction}},\ }\href
  {https://doi.org/10.1103/PhysRev.97.1474} {\bibfield  {journal} {\bibinfo
  {journal} {Phys. Rev.}\ }\textbf {\bibinfo {volume} {97}},\ \bibinfo {pages}
  {1474} (\bibinfo {year} {1955})}\BibitemShut {NoStop}%
\bibitem [{\citenamefont {Resare}\ and\ \citenamefont
  {Hofmann}(2022)}]{resare22}%
  \BibitemOpen
  \bibfield  {author} {\bibinfo {author} {\bibfnamefont {F.}~\bibnamefont
  {Resare}}\ and\ \bibinfo {author} {\bibfnamefont {J.}~\bibnamefont
  {Hofmann}},\ }\bibfield  {title} {\bibinfo {title} {{Few-to-many particle
  crossover of pair excitations in a superfluid}},\ }\bibfield  {journal}
  {\bibinfo  {journal} {arXiv preprint arXiv:2208.03762}\ }\href
  {https://doi.org/10.48550/arXiv.2208.03762} {10.48550/arXiv.2208.03762}
  (\bibinfo {year} {2022})\BibitemShut {NoStop}%
\bibitem [{\citenamefont {Zwerger}(2016)}]{zwerger16}%
  \BibitemOpen
  \bibfield  {author} {\bibinfo {author} {\bibfnamefont {W.}~\bibnamefont
  {Zwerger}},\ }\bibfield  {title} {\bibinfo {title} {{Strongly Interacting
  Fermi Gases}},\ }in\ \href {https://doi.org/10.48550/arXiv.1608.00457} {\emph
  {\bibinfo {booktitle} {Proceedings of the International School of Physics
  "Enrico Fermi" - Course 191 "Quantum Matter at Ultralow Temperatures"}}},\
  \bibinfo {editor} {edited by\ \bibinfo {editor} {\bibfnamefont
  {M.}~\bibnamefont {Inguscio}}, \bibinfo {editor} {\bibfnamefont
  {W.}~\bibnamefont {Ketterle}}, \bibinfo {editor} {\bibfnamefont
  {S.}~\bibnamefont {Stringari}},\ and\ \bibinfo {editor} {\bibfnamefont
  {G.}~\bibnamefont {Roati}}}\ (\bibinfo {address} {arXiv:1608.00457},\
  \bibinfo {year} {2016})\ p.~\bibinfo {pages} {63}\BibitemShut {NoStop}%
\bibitem [{\citenamefont {Hofmann}\ and\ \citenamefont
  {Zwerger}(2021)}]{hofmann21}%
  \BibitemOpen
  \bibfield  {author} {\bibinfo {author} {\bibfnamefont {J.}~\bibnamefont
  {Hofmann}}\ and\ \bibinfo {author} {\bibfnamefont {W.}~\bibnamefont
  {Zwerger}},\ }\bibfield  {title} {\bibinfo {title} {{Universal relations for
  dipolar quantum gases}},\ }\href
  {https://doi.org/10.1103/PhysRevResearch.3.013088} {\bibfield  {journal}
  {\bibinfo  {journal} {Phys. Rev. Research}\ }\textbf {\bibinfo {volume}
  {3}},\ \bibinfo {pages} {013088} (\bibinfo {year} {2021})}\BibitemShut
  {NoStop}%
\bibitem [{\citenamefont {Sakurai}(1994)}]{sakurai94}%
  \BibitemOpen
  \bibfield  {author} {\bibinfo {author} {\bibfnamefont {J.~J.}\ \bibnamefont
  {Sakurai}},\ }\href {https://doi.org/10.1017/9781108587280} {\emph {\bibinfo
  {title} {{Modern Quantum Mechanics}}}}\ (\bibinfo  {publisher}
  {Addison-Wesley (Reading, Massachusetts)},\ \bibinfo {year}
  {1994})\BibitemShut {NoStop}%
\bibitem [{\citenamefont {Gottfried}\ and\ \citenamefont
  {Yan}(2003)}]{gottfried03}%
  \BibitemOpen
  \bibfield  {author} {\bibinfo {author} {\bibfnamefont {K.}~\bibnamefont
  {Gottfried}}\ and\ \bibinfo {author} {\bibfnamefont {T.-M.}\ \bibnamefont
  {Yan}},\ }\href {https://doi.org/10.1007/978-0-387-21623-2} {\emph {\bibinfo
  {title} {{Quantum Mechanics: Fundamentals}}}}\ (\bibinfo  {publisher}
  {Springer (New-York)},\ \bibinfo {year} {2003})\BibitemShut {NoStop}%
\bibitem [{\citenamefont {Palm}\ \emph {et~al.}(2020)\citenamefont {Palm},
  \citenamefont {Grusdt},\ and\ \citenamefont {Preiss}}]{palm20}%
  \BibitemOpen
  \bibfield  {author} {\bibinfo {author} {\bibfnamefont {L.}~\bibnamefont
  {Palm}}, \bibinfo {author} {\bibfnamefont {F.}~\bibnamefont {Grusdt}},\ and\
  \bibinfo {author} {\bibfnamefont {P.~M.}\ \bibnamefont {Preiss}},\ }\bibfield
   {title} {\bibinfo {title} {{Skyrmion ground states of rapidly rotating
  few-fermion systems}},\ }\href {https://doi.org/10.1088/1367-2630/aba30e}
  {\bibfield  {journal} {\bibinfo  {journal} {New Journal of Physics}\ }\textbf
  {\bibinfo {volume} {22}},\ \bibinfo {pages} {083037} (\bibinfo {year}
  {2020})}\BibitemShut {NoStop}%
\bibitem [{\citenamefont {Taylor}\ and\ \citenamefont
  {Randeria}(2012)}]{taylor12}%
  \BibitemOpen
  \bibfield  {author} {\bibinfo {author} {\bibfnamefont {E.}~\bibnamefont
  {Taylor}}\ and\ \bibinfo {author} {\bibfnamefont {M.}~\bibnamefont
  {Randeria}},\ }\bibfield  {title} {\bibinfo {title} {{Apparent Low-Energy
  Scale Invariance in Two-Dimensional Fermi Gases}},\ }\href
  {https://doi.org/10.1103/PhysRevLett.109.135301} {\bibfield  {journal}
  {\bibinfo  {journal} {Phys. Rev. Lett.}\ }\textbf {\bibinfo {volume} {109}},\
  \bibinfo {pages} {135301} (\bibinfo {year} {2012})}\BibitemShut {NoStop}%
\bibitem [{\citenamefont {Mashkevich}\ \emph {et~al.}(2007)\citenamefont
  {Mashkevich}, \citenamefont {Matveenko},\ and\ \citenamefont
  {Ouvry}}]{MASHKEVICH07}%
  \BibitemOpen
  \bibfield  {author} {\bibinfo {author} {\bibfnamefont {S.}~\bibnamefont
  {Mashkevich}}, \bibinfo {author} {\bibfnamefont {S.}~\bibnamefont
  {Matveenko}},\ and\ \bibinfo {author} {\bibfnamefont {S.}~\bibnamefont
  {Ouvry}},\ }\bibfield  {title} {\bibinfo {title} {{Exact results for the
  spectra of bosons and fermions with contact interaction}},\ }\href
  {https://doi.org/https://doi.org/10.1016/j.nuclphysb.2006.11.022} {\bibfield
  {journal} {\bibinfo  {journal} {Nuclear Physics B}\ }\textbf {\bibinfo
  {volume} {763}},\ \bibinfo {pages} {431} (\bibinfo {year}
  {2007})}\BibitemShut {NoStop}%
\bibitem [{\citenamefont {Mashkevich}\ \emph {et~al.}(2013)\citenamefont
  {Mashkevich}, \citenamefont {Matveenko},\ and\ \citenamefont
  {Ouvry}}]{mashkevich11}%
  \BibitemOpen
  \bibfield  {author} {\bibinfo {author} {\bibfnamefont {S.}~\bibnamefont
  {Mashkevich}}, \bibinfo {author} {\bibfnamefont {S.}~\bibnamefont
  {Matveenko}},\ and\ \bibinfo {author} {\bibfnamefont {S.}~\bibnamefont
  {Ouvry}},\ }\bibfield  {title} {\bibinfo {title} {{Exact results for the
  spectra of interacting bosons and fermions on the lowest landau level}},\
  }\href {https://doi.org/10.1088/1742-5468/2013/02/P02013} {\bibfield
  {journal} {\bibinfo  {journal} {Journal of Statistical Mechanics: Theory and
  Experiment}\ }\textbf {\bibinfo {volume} {2013}},\ \bibinfo {pages} {P02013}
  (\bibinfo {year} {2013})}\BibitemShut {NoStop}%
\bibitem [{\citenamefont {MacDonald}(1994)}]{macdonald94}%
  \BibitemOpen
  \bibfield  {author} {\bibinfo {author} {\bibfnamefont {A.~H.}\ \bibnamefont
  {MacDonald}},\ }\href@noop {} {\bibinfo {title} {{Introduction to the Physics
  of the Quantum Hall Regime}}} (\bibinfo {year} {1994}),\ \Eprint
  {https://arxiv.org/abs/cond-mat/9410047} {arXiv:cond-mat/9410047}
  \BibitemShut {NoStop}%
\bibitem [{\citenamefont {Kohn}(1961)}]{kohn61}%
  \BibitemOpen
  \bibfield  {author} {\bibinfo {author} {\bibfnamefont {W.}~\bibnamefont
  {Kohn}},\ }\bibfield  {title} {\bibinfo {title} {{Cyclotron Resonance and de
  Haas-van Alphen Oscillations of an Interacting Electron Gas}},\ }\href
  {https://doi.org/10.1103/PhysRev.123.1242} {\bibfield  {journal} {\bibinfo
  {journal} {Phys. Rev.}\ }\textbf {\bibinfo {volume} {123}},\ \bibinfo {pages}
  {1242} (\bibinfo {year} {1961})}\BibitemShut {NoStop}%
\bibitem [{\citenamefont {Kasner}(2002)}]{kasner02}%
  \BibitemOpen
  \bibfield  {author} {\bibinfo {author} {\bibfnamefont {M.}~\bibnamefont
  {Kasner}},\ }\bibfield  {title} {\bibinfo {title} {{Electronic correlation in
  the quantum Hall regime}},\ }\href
  {https://doi.org/https://doi.org/10.1002/andp.20025140301} {\bibfield
  {journal} {\bibinfo  {journal} {Annalen der Physik}\ }\textbf {\bibinfo
  {volume} {514}},\ \bibinfo {pages} {175} (\bibinfo {year}
  {2002})}\BibitemShut {NoStop}%
\bibitem [{\citenamefont {Moroz}(2012)}]{moroz12}%
  \BibitemOpen
  \bibfield  {author} {\bibinfo {author} {\bibfnamefont {S.}~\bibnamefont
  {Moroz}},\ }\bibfield  {title} {\bibinfo {title} {{Scale-invariant Fermi gas
  in a time-dependent harmonic potential}},\ }\href
  {https://doi.org/10.1103/PhysRevA.86.011601} {\bibfield  {journal} {\bibinfo
  {journal} {Phys. Rev. A}\ }\textbf {\bibinfo {volume} {86}},\ \bibinfo
  {pages} {011601(R)} (\bibinfo {year} {2012})}\BibitemShut {NoStop}%
\bibitem [{\citenamefont {Halperin}(1983)}]{halperin83}%
  \BibitemOpen
  \bibfield  {author} {\bibinfo {author} {\bibfnamefont {B.~I.}\ \bibnamefont
  {Halperin}},\ }\bibfield  {title} {\bibinfo {title} {{Theory of the quantized
  Hall conductance}},\ }\href {https://doi.org/10.5169/seals-115362} {\bibfield
   {journal} {\bibinfo  {journal} {Helvetica Physica Acta}\ ,\ \bibinfo {pages}
  {75}} (\bibinfo {year} {1983})}\BibitemShut {NoStop}%
\bibitem [{\citenamefont {Tong}(2016)}]{tong16}%
  \BibitemOpen
  \bibfield  {author} {\bibinfo {author} {\bibfnamefont {D.}~\bibnamefont
  {Tong}},\ }\bibfield  {title} {\bibinfo {title} {{Lectures on the quantum
  Hall effect}},\ }\bibfield  {journal} {\bibinfo  {journal} {arXiv preprint
  arXiv:1606.06687}\ }\href {https://doi.org/10.48550/arXiv.1606.06687}
  {10.48550/arXiv.1606.06687} (\bibinfo {year} {2016})\BibitemShut {NoStop}%
\bibitem [{\citenamefont {Blume}\ \emph {et~al.}(2007)\citenamefont {Blume},
  \citenamefont {von Stecher},\ and\ \citenamefont {Greene}}]{blume07}%
  \BibitemOpen
  \bibfield  {author} {\bibinfo {author} {\bibfnamefont {D.}~\bibnamefont
  {Blume}}, \bibinfo {author} {\bibfnamefont {J.}~\bibnamefont {von Stecher}},\
  and\ \bibinfo {author} {\bibfnamefont {C.~H.}\ \bibnamefont {Greene}},\
  }\bibfield  {title} {\bibinfo {title} {{Universal Properties of a Trapped
  Two-Component Fermi Gas at Unitarity}},\ }\href
  {https://doi.org/10.1103/PhysRevLett.99.233201} {\bibfield  {journal}
  {\bibinfo  {journal} {Phys. Rev. Lett.}\ }\textbf {\bibinfo {volume} {99}},\
  \bibinfo {pages} {233201} (\bibinfo {year} {2007})}\BibitemShut {NoStop}%
\bibitem [{\citenamefont {Giuliani}\ and\ \citenamefont
  {Vignale}(2005)}]{giuliani05}%
  \BibitemOpen
  \bibfield  {author} {\bibinfo {author} {\bibfnamefont {G.~F.}\ \bibnamefont
  {Giuliani}}\ and\ \bibinfo {author} {\bibfnamefont {G.}~\bibnamefont
  {Vignale}},\ }\href {https://doi.org/10.1017/CBO9780511619915} {\emph
  {\bibinfo {title} {{Quantum Theory of the Electron Liquid}}}}\ (\bibinfo
  {publisher} {Cambridge University Press (Cambridge)},\ \bibinfo {year}
  {2005})\BibitemShut {NoStop}%
\bibitem [{\citenamefont {Bamler}\ and\ \citenamefont
  {Rosch}(2015)}]{bamler15}%
  \BibitemOpen
  \bibfield  {author} {\bibinfo {author} {\bibfnamefont {R.}~\bibnamefont
  {Bamler}}\ and\ \bibinfo {author} {\bibfnamefont {A.}~\bibnamefont {Rosch}},\
  }\bibfield  {title} {\bibinfo {title} {{Equilibration and approximate
  conservation laws: Dipole oscillations and perfect drag of ultracold atoms in
  a harmonic trap}},\ }\href {https://doi.org/10.1103/PhysRevA.91.063604}
  {\bibfield  {journal} {\bibinfo  {journal} {Phys. Rev. A}\ }\textbf {\bibinfo
  {volume} {91}},\ \bibinfo {pages} {063604} (\bibinfo {year}
  {2015})}\BibitemShut {NoStop}%
\bibitem [{\citenamefont {Maki}\ \emph {et~al.}(2018)\citenamefont {Maki},
  \citenamefont {Zhao},\ and\ \citenamefont {Zhou}}]{maki18}%
  \BibitemOpen
  \bibfield  {author} {\bibinfo {author} {\bibfnamefont {J.}~\bibnamefont
  {Maki}}, \bibinfo {author} {\bibfnamefont {L.-M.}\ \bibnamefont {Zhao}},\
  and\ \bibinfo {author} {\bibfnamefont {F.}~\bibnamefont {Zhou}},\ }\bibfield
  {title} {\bibinfo {title} {{Nonperturbative dynamical effects in
  nearly-scale-invariant systems: The action of breaking scale invariance}},\
  }\href {https://doi.org/10.1103/PhysRevA.98.013602} {\bibfield  {journal}
  {\bibinfo  {journal} {Phys. Rev. A}\ }\textbf {\bibinfo {volume} {98}},\
  \bibinfo {pages} {013602} (\bibinfo {year} {2018})}\BibitemShut {NoStop}%
\bibitem [{\citenamefont {Maki}\ and\ \citenamefont {Zhou}(2019)}]{maki19}%
  \BibitemOpen
  \bibfield  {author} {\bibinfo {author} {\bibfnamefont {J.}~\bibnamefont
  {Maki}}\ and\ \bibinfo {author} {\bibfnamefont {F.}~\bibnamefont {Zhou}},\
  }\bibfield  {title} {\bibinfo {title} {{Quantum many-body conformal dynamics:
  Symmetries, geometry, conformal tower states, and entropy production}},\
  }\href {https://doi.org/10.1103/PhysRevA.100.023601} {\bibfield  {journal}
  {\bibinfo  {journal} {Phys. Rev. A}\ }\textbf {\bibinfo {volume} {100}},\
  \bibinfo {pages} {023601} (\bibinfo {year} {2019})}\BibitemShut {NoStop}%
\bibitem [{\citenamefont {Saint-Jalm}\ \emph {et~al.}(2019)\citenamefont
  {Saint-Jalm}, \citenamefont {Castilho}, \citenamefont {Le~Cerf},
  \citenamefont {Bakkali-Hassani}, \citenamefont {Ville}, \citenamefont
  {Nascimbene}, \citenamefont {Beugnon},\ and\ \citenamefont
  {Dalibard}}]{saintjalm19}%
  \BibitemOpen
  \bibfield  {author} {\bibinfo {author} {\bibfnamefont {R.}~\bibnamefont
  {Saint-Jalm}}, \bibinfo {author} {\bibfnamefont {P.~C.~M.}\ \bibnamefont
  {Castilho}}, \bibinfo {author} {\bibfnamefont {E.}~\bibnamefont {Le~Cerf}},
  \bibinfo {author} {\bibfnamefont {B.}~\bibnamefont {Bakkali-Hassani}},
  \bibinfo {author} {\bibfnamefont {J.-L.}\ \bibnamefont {Ville}}, \bibinfo
  {author} {\bibfnamefont {S.}~\bibnamefont {Nascimbene}}, \bibinfo {author}
  {\bibfnamefont {J.}~\bibnamefont {Beugnon}},\ and\ \bibinfo {author}
  {\bibfnamefont {J.}~\bibnamefont {Dalibard}},\ }\bibfield  {title} {\bibinfo
  {title} {{Dynamical Symmetry and Breathers in a Two-Dimensional Bose Gas}},\
  }\href {https://doi.org/10.1103/PhysRevX.9.021035} {\bibfield  {journal}
  {\bibinfo  {journal} {Phys. Rev. X}\ }\textbf {\bibinfo {volume} {9}},\
  \bibinfo {pages} {021035} (\bibinfo {year} {2019})}\BibitemShut {NoStop}%
\bibitem [{\citenamefont {Lv}\ \emph {et~al.}(2020)\citenamefont {Lv},
  \citenamefont {Zhang},\ and\ \citenamefont {Zhou}}]{lv20}%
  \BibitemOpen
  \bibfield  {author} {\bibinfo {author} {\bibfnamefont {C.}~\bibnamefont
  {Lv}}, \bibinfo {author} {\bibfnamefont {R.}~\bibnamefont {Zhang}},\ and\
  \bibinfo {author} {\bibfnamefont {Q.}~\bibnamefont {Zhou}},\ }\bibfield
  {title} {\bibinfo {title} {{$SU(1,1)$ Echoes for Breathers in Quantum
  Gases}},\ }\href {https://doi.org/10.1103/PhysRevLett.125.253002} {\bibfield
  {journal} {\bibinfo  {journal} {Phys. Rev. Lett.}\ }\textbf {\bibinfo
  {volume} {125}},\ \bibinfo {pages} {253002} (\bibinfo {year}
  {2020})}\BibitemShut {NoStop}%
\bibitem [{\citenamefont {Shi}\ \emph {et~al.}(2021)\citenamefont {Shi},
  \citenamefont {Gao},\ and\ \citenamefont {Zhai}}]{shi20}%
  \BibitemOpen
  \bibfield  {author} {\bibinfo {author} {\bibfnamefont {Z.-Y.}\ \bibnamefont
  {Shi}}, \bibinfo {author} {\bibfnamefont {C.}~\bibnamefont {Gao}},\ and\
  \bibinfo {author} {\bibfnamefont {H.}~\bibnamefont {Zhai}},\ }\bibfield
  {title} {\bibinfo {title} {{Ideal-Gas Approach to Hydrodynamics}},\ }\href
  {https://doi.org/10.1103/PhysRevX.11.041031} {\bibfield  {journal} {\bibinfo
  {journal} {Phys. Rev. X}\ }\textbf {\bibinfo {volume} {11}},\ \bibinfo
  {pages} {041031} (\bibinfo {year} {2021})}\BibitemShut {NoStop}%
\bibitem [{\citenamefont {Olshanii}\ \emph {et~al.}(2021)\citenamefont
  {Olshanii}, \citenamefont {Deshommes}, \citenamefont {Torrents},
  \citenamefont {Gonchenko}, \citenamefont {Dunjko},\ and\ \citenamefont
  {Astrakharchik}}]{olshanii21}%
  \BibitemOpen
  \bibfield  {author} {\bibinfo {author} {\bibfnamefont {M.}~\bibnamefont
  {Olshanii}}, \bibinfo {author} {\bibfnamefont {D.}~\bibnamefont {Deshommes}},
  \bibinfo {author} {\bibfnamefont {J.}~\bibnamefont {Torrents}}, \bibinfo
  {author} {\bibfnamefont {M.}~\bibnamefont {Gonchenko}}, \bibinfo {author}
  {\bibfnamefont {V.}~\bibnamefont {Dunjko}},\ and\ \bibinfo {author}
  {\bibfnamefont {G.~E.}\ \bibnamefont {Astrakharchik}},\ }\bibfield  {title}
  {\bibinfo {title} {{Triangular Gross-Pitaevskii breathers and
  Damski-Chandrasekhar shock waves}},\ }\href
  {https://doi.org/10.21468/SciPostPhys.10.5.114} {\bibfield  {journal}
  {\bibinfo  {journal} {SciPost Phys.}\ }\textbf {\bibinfo {volume} {10}},\
  \bibinfo {pages} {114} (\bibinfo {year} {2021})}\BibitemShut {NoStop}%
\bibitem [{\citenamefont {Maki}\ and\ \citenamefont
  {Zhou}(2024)}]{maki2023emergent}%
  \BibitemOpen
  \bibfield  {author} {\bibinfo {author} {\bibfnamefont {J.}~\bibnamefont
  {Maki}}\ and\ \bibinfo {author} {\bibfnamefont {F.}~\bibnamefont {Zhou}},\
  }\bibfield  {title} {\bibinfo {title} {Emergent infrared conformal dynamics:
  Applications to strongly interacting quantum states},\ }\href
  {https://doi.org/10.1103/PhysRevA.109.L051303} {\bibfield  {journal}
  {\bibinfo  {journal} {Phys. Rev. A}\ }\textbf {\bibinfo {volume} {109}},\
  \bibinfo {pages} {L051303} (\bibinfo {year} {2024})}\BibitemShut {NoStop}%
\bibitem [{\citenamefont {Lunt}\ \emph {et~al.}(2024)\citenamefont {Lunt},
  \citenamefont {Hill}, \citenamefont {Reiter}, \citenamefont {Preiss},
  \citenamefont {Gałka},\ and\ \citenamefont {Jochim}}]{lunt2024}%
  \BibitemOpen
  \bibfield  {author} {\bibinfo {author} {\bibfnamefont {P.}~\bibnamefont
  {Lunt}}, \bibinfo {author} {\bibfnamefont {P.}~\bibnamefont {Hill}}, \bibinfo
  {author} {\bibfnamefont {J.}~\bibnamefont {Reiter}}, \bibinfo {author}
  {\bibfnamefont {P.~M.}\ \bibnamefont {Preiss}}, \bibinfo {author}
  {\bibfnamefont {M.}~\bibnamefont {Gałka}},\ and\ \bibinfo {author}
  {\bibfnamefont {S.}~\bibnamefont {Jochim}},\ }\bibfield  {title} {\bibinfo
  {title} {{Realization of a Laughlin state of two rapidly rotating
  fermions}},\ }\bibfield  {journal} {\bibinfo  {journal} {arXiv preprint
  arXiv:2402.14814}\ }\href {https://doi.org/10.48550/arXiv.2402.14814}
  {10.48550/arXiv.2402.14814} (\bibinfo {year} {2024})\BibitemShut {NoStop}%
\end{thebibliography}%

\end{document}